\documentclass{amsart}
\usepackage{amssymb}
\usepackage{epsfig}

\begin{document}
\title{Atom-Photon Interactions with Respect to Quantum Computation:
Three-Level Atom in Two-Mode Field}
\author{G.~K.~Giorgadze}
\address{
Joint Institute for Nuclear Research, Dubna, Russia}
\email{ggk@jinr.ru}
\author{Z.~G.~Melikishvili}
\address{Institute of Cybernetics,
         Georgian Academy of Sciences,
         Tbilisi, Georgia}%
\email{z\_melikishvili@posta.ge}%

\thanks{To appear in ``Contemporary mathematics and its
applications''}

 \maketitle
\bibliographystyle{plain}
\begin{abstract}
In the paper, analysis of a quantum optical system---three-level
atom in a quantum electromagnetic field is given. Evolution
operators  are constructed in closed form.
\end{abstract}

\section{Introduction}
In theory of computation it is known that the three-state element is
most efficient for classical computers, but its realization remained
a problem for many years. The progress was achieved in 90-ties (see
\cite{liu}, \cite{zhang}, \cite{arestova}). Optimal number of states
of the physical system for the expression  mod $p$, where $p$ is a
prime, is 3. Indeed, if we denote by $N_p$ this number, it is known
that $N_p=f(p)N_2,$ where $f(p)=\frac{p}{2\log_2p},$ we obtain that
the function $\frac{N_p}{N_2}$ have the minimum when $p=3.$ The
necessary mathematical theory in classical many-valued logic is
given in \cite{iab}.

In the quantum world the many-level systems are ordinary ones. Thus
in this paper  we will consider one possible version of a quantum
processor based on a many-level quantum system.

In \cite{erco} the extension of the universal quantum logic to the
multi-valued domain is considered, where the unit of memory is the
qudit. Thereby it is assumed that arbitrary unitary operators on
any $d$-level system, where $d>2$ is any number, can be decomposed
into logical gates that operate on only two levels at a time. For
example, the linear ion trap quantum computer \cite{cirac} uses
only two levels in each ion for computing although additional
levels can be accessed, and are typically needed, for processing
and reading out the state of the ion (see \cite{erco} and the
literature cited there).

In the case when $d=3$, the set of universal gates is built in the
following manner.

Let us introduce the following unitary operators on $C^3\rightarrow
C^3:$
$$
X_3(\phi)=\left (
\begin{array}{cccc}
1 & 0 & 0\\
0&1&0\\
0&0&e^{i\phi}
\end{array}
\right), Z_3(c_0,c_1,c_2)=\left (
\begin{array}{cccc}
c_0 & 0 & -c_2\\
0&-c_1&c_2\\
-\overline{c}_2&\overline{c}_2&\overline{c}_2
\end{array}
\right).
$$
It can be easily checked that $X_3(\phi)$ and $Z_3(c_0,c_1,c_2)$
act on the states $|0>,$ $|1>,$ $|2>$ as
$$
X_3(\phi)|0>=|0>, X_3(\phi)|1>=|1>, X_3(\phi)|2>=e^{i\phi}|2>;
$$
$$
Z_3(\phi)|0>=c_0|0>-\overline{c}_2|2>,
Z_3(\phi)|1>=-c_1|0>+\overline{c}_2|2>,
Z_3|2>=-c_2|0>+c_2|1>+\overline{c}_2|2>.
$$
Furthermore, the gate $Z_3(c_0,c_1,c_2)$ satisfies the following
condition:
$$
Z_3(c_0,c_1,c_2)=\left (
\begin{array}{cccc}
c_0 & 0 & -c_2\\
0&-c_1&c_2\\
-\overline{c}_2&\overline{c}_2&\overline{c}_2
\end{array}
\right)
(c_0|0>+c_1|1>+c_2|2>)=
$$
$$
=(-c_2c_0+c_0c_2,-c_2c_1+c_1c_2,|c_0|^2+|c_1|^2+|c_2|^2=|2>).
$$
In \cite{erco} it is proved that  the set of gates $\{ X_3(\phi),
Z_3(c_0,c_1,c_2), G_2[X_3(\phi)], G_2[Z_3(c_0,c)1,c_2)] \}$ is
universal for a three-level quantum system, where $G_2[Y]=\left (
\begin{array}{cccc}
I & 0 \\
0&Y
\end{array}
\right)$ and $I$ is the identity matrix.

Below is given another approach of description of quantum system
which gives the single qutrit gates, in particular, we obtain in
closed form evolution operators depending on parameters.

\section{Atom-Photon Interactions Hamiltonian}

We will build the special case Hamiltonian of a hydrogen-like atom
moving and interacting with a quantized electromagnetic field, at
that using the notations of \cite{shleikh} and \cite{puri}. The
general form of such Hamiltonian, including the center-of-mass
motion and all the modes of the interacting field, is presented in
\cite{shleikh}.  For our purposes we consider the following
conditions on atom-photon interactions: (i) All the atom-photon
interactions are the electric dipole; (ii) Only three atomic
levels are included in the interaction; (iii) Two quantized
resonator modes are interacting with this three-level system; (iv)
Each of these modes interacts with only one couple of levels; (v)
So, from three possible couples of levels, only two of them
interact directly. Levels of the third couple interact only by
means of an intermediate level; (vi) In this work we permit, but
do not consider the center-of-mass motion and large detuning from
resonances, which cause the state entanglement.

Let us consider atom-photon interactions Hamiltonian in the
electric dipole approximation
\begin{equation}\label{1}
\hat{H}_{\mathbf{r}\centerdot\mathbf{E}}=-\wp
\mathbf{E}(R,t)=-e\mathbf{r}\mathbf{E}(\mathbf{R},t),
\end{equation}
and transform it for three-level atom interacting with two modes
of quantized fields of radiation. For these purposes previously we
must discuss the following operators.

\textbf{Hamiltonian for internal degrees of freedom of
non-interacting atom.} We consider internal degrees of freedom of a
three-level atom. Due to assumptions (i-vi) only two couples of
energy states are involved in interaction. In all cases the
intermediate state with energy $E_i\equiv\hbar\omega_i$  is labeled
by the state vector $|i>$. From other states, the state with higher
energy is named excited, denoted $E_e\equiv\hbar\omega_e$  and
labeled by the state vector $|e>$. The remaining energy level is
named ground level, denoted $E_g\equiv\hbar\omega_g$ and labeled by
the state vector $|g>$ (see Fig.1). All of them are eigenstates of
the Hamiltonian $\hat{H}_{atom}$ of non-interacting three-level
atom:
\begin{equation}\label{2}
\hat{H}_{atom}|e>=\hbar\omega_e|e>, \text{  }
\hat{H}_{atom}|i>=\hbar\omega_i|i>, \text{  }
\hat{H}_{atom}|g>=\hbar\omega_g|g>.
\end{equation}
Using the completeness relation $|e><e|+|i><i|+|g><g|=1,$ we can
write the arbitrary atomic operator $\hat{\mathbf{O}}$  as
\begin{equation}\label{3}
\hat{\mathbf{O}}=1 \hat{\mathbf{O}}
1=\sum_{j,j^{\prime}=e,g,i}|j><j|\hat{\mathbf{O}}|j^{\prime}><j^{\prime}|.
\end{equation}
Utilizing (\ref{3}) and the orthonormality condition
$<j||j^{\prime}>=\delta_{ij^{\prime}}$ for the Hamiltonian of a
three-level atom, one can easily obtain
\begin{equation}\label{4}
\hat{H}_{atom}=(|e><e|+|i><i|+|g><g|)\hat{H}_{atom}(|e><e|+|i><i|+|g><g|)=
\end{equation}
$$
E_e|e><e|+E_i|i><i|+E_g|g><g|,
$$
where we use (\ref{2}) and above notations $E_e=\hbar\omega_e$,
$E_i=\hbar\omega_i$, $E_g=\hbar\omega_g$.  Then
$$
|e>=\left (
\begin{array}{cccc}
1 \\
0\\
0
\end{array}
\right), <e|=(1,0,0), |i>=\left (
\begin{array}{cccc}
0 \\
1\\
0
\end{array}
\right), <i|=(0,1,0), |g>=\left (
\begin{array}{cccc}
0 \\
0\\
1
\end{array}
\right), <g|=(0,0,1),
$$
Thus, the Hamiltonian (\ref{4}) has the following matrix form
\begin{equation}\label{6}
\hat{H}_{atom}=E_e\left (
\begin{array}{cccc}
1 & 0 & 0 \\
0 & 0 & 0 \\
0 & 0 & 0
\end{array}
\right ) + E_i\left (
\begin{array}{cccc}
0 & 0 & 0 \\
0 & 1 & 0 \\
0 & 0 & 0
\end{array}
\right ) + E_g\left (
\begin{array}{cccc}
0 & 0 & 0 \\
0 & 0 & 0 \\
0 & 0 & 1
\end{array}
\right )=
\end{equation}
$$
=\left (
\begin{array}{cccc}
E_g & 0 & 0 \\
0 & E_i & 0 \\
0 & 0 & E_g
\end{array}
\right ) =\hbar\left (
\begin{array}{cccc}
\omega_e & 0 & 0 \\
0 & \omega_i & 0 \\
0 & 0 & \omega_g
\end{array}
\right ).
$$

\

\textbf{The dipole moment operator.} Now we can use the completeness
relation once more to express the position operator
$\hat{\mathbf{r}}$ in terms of energy eigenstates. As each state,
wave functions $\Psi_j(\mathbf{r})$ of a free atom have definite
parity, then the diagonal matrix elements equal zero
$$
<j|\hat{\mathbf{r}}|j>=\int|\Psi_j(\mathbf{r})|^2\mathbf{r}d^3r=0.
$$
Indeed, if the function $|\Psi_j(\mathbf{r})|^2$  is symmetric and
the operator $\mathbf{r}$ antisymmetric, the expression under the
integral sign is antisymmetric. For the nondiagonal matrix elements
we have the following expressions
\begin{equation}\label{7}
e<i|\hat{\mathbf{r}}|g>=e\int
\Psi_i^{*}(\mathbf{r})\mathbf{r}\Psi_g(\mathbf{r})d^3r=\wp_{ig},
\end{equation}
and
$$
e<g|\hat{\mathbf{r}}|i>=e\int
\Psi_g^{*}(\mathbf{r})\mathbf{r}\Psi_i(\mathbf{r})d^3r=\wp_{ig}^{*}=\wp_{gi}.
$$
In the same way
$$
e<e|\hat{\mathbf{r}}|i>=e\int
\Psi_i^{*}(\mathbf{r})\mathbf{r}\Psi_g(\mathbf{r})d^3r=\wp_{ei},
$$
and
$$
e<i|\hat{\mathbf{r}}|e>=e\int
\Psi_g^{*}(\mathbf{r})\mathbf{r}\Psi_i(\mathbf{r})d^3r=\wp_{ei}^{*}=\wp_{ie}.
$$
Thus, the dipole moment operator $e\hat{\mathbf{r}}$ is expressed in
the following manner
\begin{equation}\label{do}
e\hat{\mathbf{r}}=\wp_{ig}|i><g|+\wp_{gi}|g><i|+\wp_{ei}|e><i|+\wp_{ie}|i><e|.
\end{equation}
Note that this operator describes transitions from the ground
state $|g>$ to the intermediate state $|i>$, from intermediate to
the excited state $|e>$, and the process in reverse order back to
$|g>$. For verification let uss apply the operator
$e\hat{\mathbf{r}}$ (9) to the ground state $|g>.$  Then we get
$$e\hat{\mathbf{r}}|g>=\wp_{ig}|i><g||g>+\wp_{gi}|g><i||g>+\wp_{ei}|e><i||g>+\wp_{ie}|i><e||g>=\wp_{ig}|i>.$$
Similarly
$$e\hat{\mathbf{r}}|e>=\wp_{ig}|i><g||e>+\wp_{gi}|g><i||e>+\wp_{ei}|e><i||e>+\wp_{ie}|i><e||e>=\wp_{ie}|i>$$
and finally
$$e\hat{\mathbf{r}}|i>=\wp_{ig}|i><g||i>+\wp_{gi}|g><i||i>+\wp_{ei}|e><i||i>+\wp_{ie}|i><e||g>=\wp_{gi}|g>+\wp_{ei}|e>.$$
Thus, the operator
\begin{equation}\label{i1}
 \hat{\sigma}_{ig}\equiv|i><g|, \text{
} |i><g|=\left (
\begin{array}{cccc}
0 \\
1 \\
0
\end{array}
\right )(0,0,1)=\left (
\begin{array}{cccc}
0 & 0 & 0 \\
0 & 0 & 1 \\
0 & 0 & 0
\end{array}
\right ),
\end{equation}
causes the atom transition from the ground to the intermediate state
and can be interpreted as a creation operator for the atom in the
intermediate state $|i>$.

On the contrary, the operator
 \begin{equation}\label{i2}
\hat{\sigma}_{gi}\equiv|g><i|, \text{  } |g><i|=\left (
\begin{array}{cccc}
0 \\
0 \\
1
\end{array}
\right )(0,1,0)=\left (
\begin{array}{cccc}
0 & 0 & 0 \\
0 & 0 & 0 \\
0 & 1 & 0
\end{array}
\right ),
\end{equation}
annihilates the atom in the intermediate state and so represents the
annihilation operator for the intermediate state.

The operator
\begin{equation}\label{i3}
\hat{\sigma}_{ei}\equiv|e><i|, \text{  } |e><i|=\left (
\begin{array}{cccc}
1 \\
0 \\
0
\end{array}
\right )(0,1,0)=\left (
\begin{array}{cccc}
0 & 1 & 0 \\
0 & 0 & 0 \\
0 & 0 & 0
\end{array}
\right ),
\end{equation}
causes transition of the atom from the intermediate to the excited
state and can be interpreted as a creation operator for the atom in
the excited state $|e>$.

The operator
\begin{equation}\label{i4}
\hat{\sigma}_{ei}\equiv|i><e|, \text{  } |i><e|=\left (
\begin{array}{cccc}
0 \\
1 \\
0
\end{array}
\right )(1,0,0)=\left (
\begin{array}{cccc}
0 & 0 & 0 \\
1 & 0 & 0 \\
0 & 0 & 0
\end{array}
\right ),
\end{equation}
annihilates the atom in the excited state and so represents the
annihilation operator for the excited state. The dipole moment
operator (\ref{do}) is defined in accordance to the expressions
(\ref{i1})-(\ref{i4}), of creation and annihilation operators, to be
\begin{equation}\label{do1}e\hat{\mathbf{r}}=\wp_{ig}\hat{\sigma}_{ig}+\wp_{gi}\hat{\sigma}_{ig}+
\wp_{ei}\hat{\sigma}_{ei}+\wp_{ie}\hat{\sigma}_{ie}.\end{equation}

\

\textbf{The electric field operator.} Now let us discuss two modes
of radiation field. We assume for determinacy, that the radiation
field is formed in three-mirror standing wave resonator and
decompose the electric field strength in the resonator as follows
\begin{equation}\label{efo}
\hat{\mathbf{E}}(\hat{\mathbf{R}},t)=\mathcal{F}_{0a}\mathbf{u}_a(\hat{\mathbf{R}})i(\hat{a}-\hat{a}^{+})+\mathcal{F}_{0b}
\mathbf{u}_b(\hat{\mathbf{R}})i(\hat{b}-\hat{b}^{+}),
\end{equation} where
$\mathcal{F}_{0j}\equiv\sqrt{\frac{\hbar\Omega_j}{2\epsilon_0V_j}},$
$j=a,b.$

Here $\mathbf{u}_j(\hat{\mathbf{R}})$  represents the mode functions
of the resonator at the position $\mathbf{R}$ of the atom.
$\Omega_a$ and $\Omega_b$ are oscillation frequencies of the field.
In a classical treatment, $\hat{a}^+,$ $\hat{b}^+$ and $\hat{a},$
$\hat{b}$ are time dependent amplitudes. In our present treatment
they are operators which obey the commutation relations
\begin{equation}\label{comm}
\hat{a}\hat{a}^+-\hat{a}^+\hat{a}=1, \text{  }
\hat{b}\hat{b}^+-\hat{b}^+\hat{b}=1.
\end{equation}
With their aid the quantized field Hamiltonian can be written in the
form
\begin{equation}\label{qfh}
\hat{H}_{field}=\hbar\Omega_a\hat{a}^+\hat{a}+\hbar\Omega_b\hat{b}^+\hat{b}.\end{equation}
Here we neglect the zero-point energy.

\

\textbf{Back to the interaction Hamiltonian.} Now consider the
 Hamiltonian of electric dipole atom-photon interactions again
(\ref{1})
 $$
 \hat{H}_{\mathbf{r}\mathbf{E}}=-e\hat{\mathbf{r}}\hat{\mathbf{E}}(\hat{\mathbf{R}},t).
 $$
Substituting in this expression the expressions of the dipole moment
operator $e\hat{\mathbf{r}}$ (\ref{do1}) and electric field operator
$\hat{\mathbf{r}}\hat{\mathbf{E}}(\hat{\mathbf{R}},t)$ we obtain
\begin{equation}\label{ih}
\hat{H}_{\mathbf{r}\mathbf{E}}=-\mathcal{F}_{0a}i[\wp_{ig}\mathbf{u}_a(\hat{\mathbf{R}})\hat{\sigma}_{ig}+
\wp_{ei}\mathbf{u}_b(\hat{\mathbf{R}})\hat{\sigma}_{gi}](\hat{a}-\hat{a}^{+})-
\end{equation}
$$
-\mathcal{F}_{0b}i[\wp_{ei}\mathbf{u}_b(\hat{\mathbf{R}})\hat{\sigma}_{ei}+
\wp_{ei}\mathbf{u}_b(\hat{\mathbf{R}})\hat{\sigma}_{ie}](\hat{b}-\hat{b}^{+}).
$$
This expression contains the scalar product $\wp\mathbf{u}$  of the
dipole moment and the mode function $\mathbf{u}$. As this product is
a complex quantity, we represent it in the following manner
\begin{equation}\label{phase}
\wp\mathbf{u}=|\wp\mathbf{u}|e^{i\varphi},
\end{equation}
where $\varphi$  is the phase. Substituting (\ref{phase}) in
(\ref{ih}) the Hamiltonian of electric dipole atom-photon
interactions can be written in the following form
\begin{equation}\label{ih1}
\hat{H}_{\mathbf{r}\mathbf{E}}=-\hbar\mathcal{F}_{0a}i[\frac{|\wp_{ig}\mathbf{u}_a(\hat{\mathbf{R}})|}{\hbar}\hat{\sigma}_{ig}e^{i\varphi_a}+
\frac{|\wp_{ei}\mathbf{u}_b(\hat{\mathbf{R}})|}{\hbar}\hat{\sigma}_{gi}e^{-i\varphi_a}](\hat{a}-\hat{a}^{+})-
\end{equation}
$$
-\hbar\mathcal{F}_{0b}i[\frac{|\wp_{ei}\mathbf{u}_b(\hat{\mathbf{R}})|}{\hbar}\hat{\sigma}_{ei}e^{i\varphi_b}+
\frac{|\wp_{ei}\mathbf{u}_b(\hat{\mathbf{R}})|}{\hbar}\hat{\sigma}_{ie}e^{-i\varphi_b}](\hat{b}-\hat{b}^{+}).
$$
Introducing the Rabi frequencies
$$
g_a(\mathbf{R})=\frac{|\wp_{ig}\mathbf{u}_a(\hat{\mathbf{R}})|}{\hbar}\mathcal{F}_{0a},
\text{   }
g_b(\mathbf{R})=\frac{|\wp_{ei}\mathbf{u}_b(\hat{\mathbf{R}})|}{\hbar}\mathcal{F}_{0b}
$$
the interaction Hamiltonian (\ref{ih1}) now is given by
\begin{equation}\label{ih2}
\hat{H}_{\mathbf{r}\mathbf{E}}=\hbar
g_a(\hat{\mathbf{R}})(-i)(\hat{\sigma}_{ig}e^{i\varphi_a}+\hat{\sigma}_{gi}e^{-i\varphi_a})(\hat{a}-\hat{a}^+)+
\end{equation}
$$
\hbar
g_b(\hat{\mathbf{R}})(-i)(\hat{\sigma}_{ei}e^{i\varphi_b}+\hat{\sigma}_{ie}e^{-i\varphi_b})(\hat{b}-\hat{b}^+).
$$
Introducing the difference of phases
$\delta\varphi\equiv\varphi_a-\varphi_b,$ in the Hamiltonian
(\ref{ih2}) we obtain
\begin{equation}\label{ih3}
\hat{H}_{\mathbf{r}\mathbf{E}}=\hbar
g_a(\hat{\mathbf{R}})(-i)(\hat{\sigma}_{ig}e^{i\varphi_a}+\hat{\sigma}_{gi}e^{-i\varphi_a})(\hat{a}-\hat{a}^+)+
\end{equation}
$$
\hbar
g_b(\hat{\mathbf{R}})(-i)(\hat{\sigma}_{ei}e^{i(\delta\varphi-\varphi_a)}+\hat{\sigma}_{ie}e^{-i(\delta\varphi-\varphi_a)})(\hat{b}-\hat{b}^+).
$$
In quantum mechanics only the difference of phases is important, so
in (\ref{ih3}) we can put $\varphi_a=\pi/2$ and finally obtain
\begin{equation}\label{ih4}
\hat{H}_{\mathbf{r}\mathbf{E}}=\hbar
g_a(\hat{\mathbf{R}})(\hat{\sigma}_{ig}-\hat{\sigma}_{gi})(\hat{a}-\hat{a}^+)+
\hbar
g_b(\hat{\mathbf{R}})(\hat{\sigma}_{ei}e^{-i\delta\varphi}-\hat{\sigma}_{ie}e^{i\delta\varphi})(\hat{b}-\hat{b}^+).
\end{equation}
Expression (\ref{ih4}) is the final form of the exact Hamiltonian
responsible for all atom-photon interactions between the three-level
atom and two modes of electromagnetic field. Just this Hamiltonian
will be used hereinafter.

\

\textbf{Total Hamiltonian.} Joining up the above main results
(\ref{4}), (\ref{qfh}), (\ref{ih4}), and taking into account, that
$$
\hat{\sigma}_{ee}\equiv|e><e|, \text{
}\hat{\sigma}_{ii}\equiv|i><i|, \text{  }
\hat{\sigma}_{gg}\equiv|g><g|,
$$
one can build the total Hamiltonian of the whole atom-field system
\begin{equation}\label{total}
\hat{H}=\frac{\hat{\mathbf{P}}^2}{2M}+\hat{H}_{field}+\hat{H}_{atom}+\hat{H}_{\mathbf{r}\mathbf{E}}=
\end{equation}
$$
\frac{\hat{\mathbf{P}}^2}{2M}+\hbar\Omega_a\hat{a}^+\hat{a}+\hbar\Omega_b\hat{b}^+\hat{b}+E_e\hat{\sigma}_{ee}
+E_i\hat{\sigma}_{ii}+E_g\hat{\sigma}_{gg}+
$$
$$
+\hbar
g_a(\hat{\mathbf{R}})(\hat{\sigma}_{ig}-\hat{\sigma}_{gi})(\hat{a}-\hat{a}^+)+
\hbar
g_b(\hat{\mathbf{R}})(\hat{\sigma}_{ei}e^{-i\delta\varphi}-\hat{\sigma}_{ie}e^{i\delta\varphi})(\hat{b}-\hat{b}^+).
$$

This Hamiltonian includes: the operator of kinetic energy
$\frac{\hat{\mathbf{P}}^2}{2M}$  of the center of mass movement;
free-field Hamiltonian
$\hat{H}_{field}=\hbar\Omega_a\hat{a}^+\hat{a}+\hbar\Omega_b\hat{b}^+\hat{b};$
the Hamiltonian responsible for internal states of the atom
$\hat{H}_{atom}=E_e\hat{\sigma}_{ee}
+E_i\hat{\sigma}_{ii}+E_g\hat{\sigma}_{gg}$  and interaction
operator $\hat{H}_{\mathbf{r}\mathbf{E}}=\hbar
g_a(\hat{\mathbf{R}})(\hat{\sigma}_{ig}-\hat{\sigma}_{gi})(\hat{a}-\hat{a}^+)+
\hbar
g_b(\hat{\mathbf{R}})(\hat{\sigma}_{ei}e^{-i\delta\varphi}-\hat{\sigma}_{ie}e^{i\delta\varphi})(\hat{b}-\hat{b}^+).
$ Connecting the internal states with field operators
$\hat{a}-\hat{a}^+,$ $\hat{b}-\hat{b}^+$ by means of atom-field
coupling constants  $g_a(\hat{\mathbf{R}}),$
$g_a(\hat{\mathbf{R}}).$

\

\section{Simplification of Atom-Photon Interactions Hamiltonian by Selecting the Real Transitions (SRT) and Rotating Wave Approximation (RWA)}
There are two reasons for simplification of the atom-photon
interactions Hamiltonian (\ref{ih4})
\begin{equation}\label{ih5}
\hat{H}_{\mathbf{r}\mathbf{E}}=\hbar
g_a(\hat{\mathbf{R}})(\hat{\sigma}_{ig}\hat{a}^++\hat{\sigma}_{gi}\hat{a}-\hat{\sigma}_{ig}\hat{a}-\hat{\sigma}_{ig}\hat{a}^+)+
\end{equation}
$$
+\hbar g_b(\hat{\mathbf{R}})(\hat{\sigma}_{ei}\hat{b}^+
e^{i\delta\varphi}-\hat{\sigma}_{ie}\hat{b}e^{-i\delta\varphi}-\hat{\sigma}_{ie}\hat{b}e^{i\delta\varphi}-\hat{\sigma}_{ie}\hat{b}^+e^{-i\delta\varphi}).
$$
Namely, it amounts to confining our analysis to real transitions in
which a photon is emitted while the atom goes from its upper state
to its lower state, or a photon is absorbed while the atom goes from
its lower state to its upper state. The second reason is
mathematical and is connected with the averaging procedure --
neglecting the fast oscillating terms. Hereinafter we will use the
interaction representation (Dirac representation) which is defined
by the internal atomic states and the free-field states and under
these assumptions we may reduce the Hamiltonian (\ref{ih5}) to the
form which is called Rotating Wave Approximation (RWA).

In a three-level system the simplification procedure of interaction
Hamiltonian should be carried out separately for each transition
scheme. Thus, first of all let us consider these transition schemes.
In the three-level atom due to (i)-(v) of section 2 there are three
different configurations possible for transitions between the three
levels depending upon the position of the intermediate level $|i>$.
Those are shown in Fig. 1 along with the allowed transitions. In the
configuration of the first diagrams in Fig. 1, the energy of $|i>$
is intermediate between the energies of the other two levels. It is
called the ladder configuration. Notation 'L' stands for this
configuration. The energy of $|i>$  in the $\Lambda$ or Raman
configuration is higher than that of the other two levels (second
diagram in Fig. 1) whereas the level $|i>$ in the V-system is below
the other two levels (third diagram in Fig. 1).

\begin{center}

\psfig{file=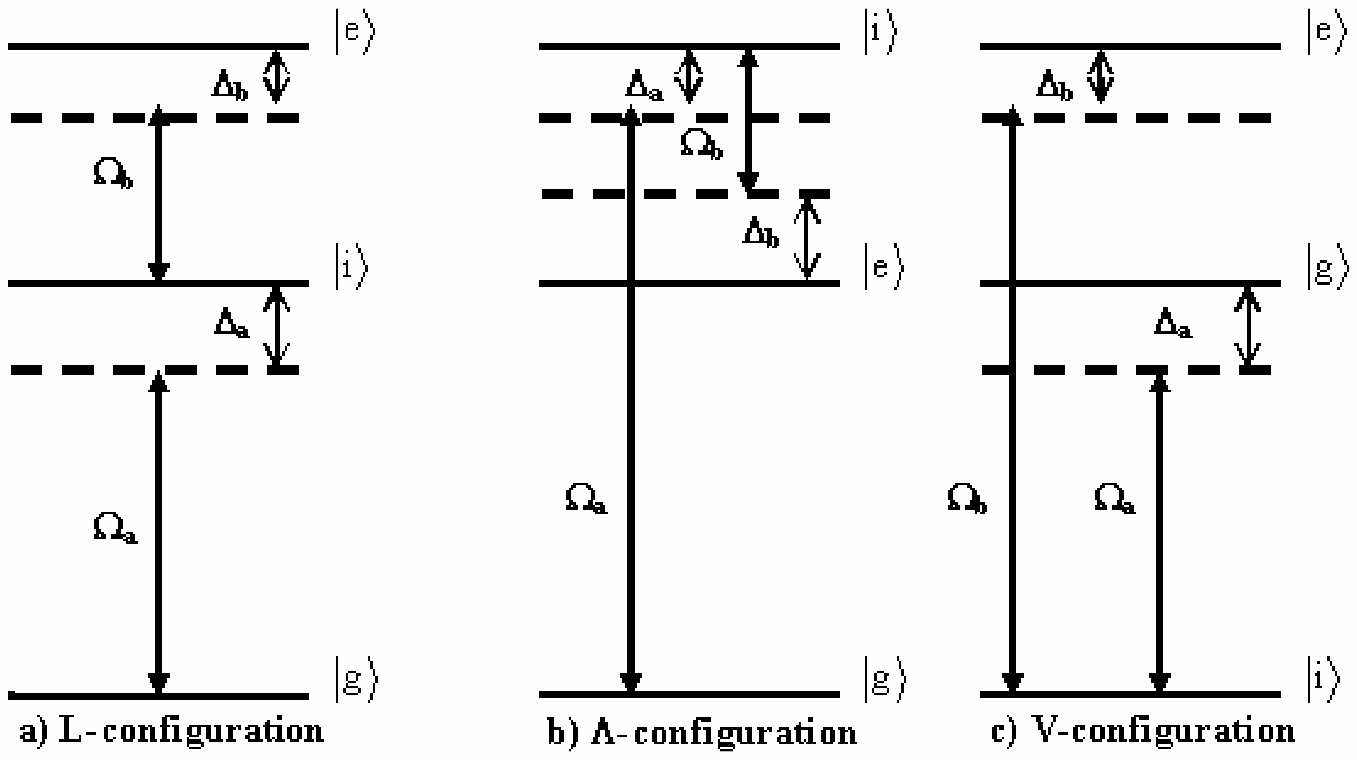}

\end{center}
\begin{center}
Figure 1. Three configuration of the three level atom in a two-mode \\
 field each acting on only one transition.
\end{center}

\subsection{Simplification for L-configuration - SRT}
\textbf{Selection of real transitions.} The terms
$\hat{\sigma}_{gi}\hat{a},$ $\hat{\sigma}_{ig}\hat{a}^+,$
$\hat{\sigma}_{ie}\hat{b}$ and $\hat{\sigma}_{ei}\hat{b}^{+},$ in
the interaction Hamiltonian (\ref{ih5}) lead to the violation of the
energy conservation law. Indeed, note that the operators
$\hat{\sigma}_{gi}\hat{a}$  and $\hat{\sigma}_{ie}\hat{b},$
annihilate the excited atom (in $|i>$ and $|e>$) and simultaneously
annihilate the field excitation (Fig. 2c). In the same manner
operators  $\hat{\sigma}_{ig}\hat{a}^+$ and
$\hat{\sigma}_{ei}\hat{b}^{+}$ create the field excitation while the
atom is annihilated in $|g>$ ($|i>$) and created in $|i>$ ($|e>$)
states (Fig. 2d).
\begin{center}
\psfig{file=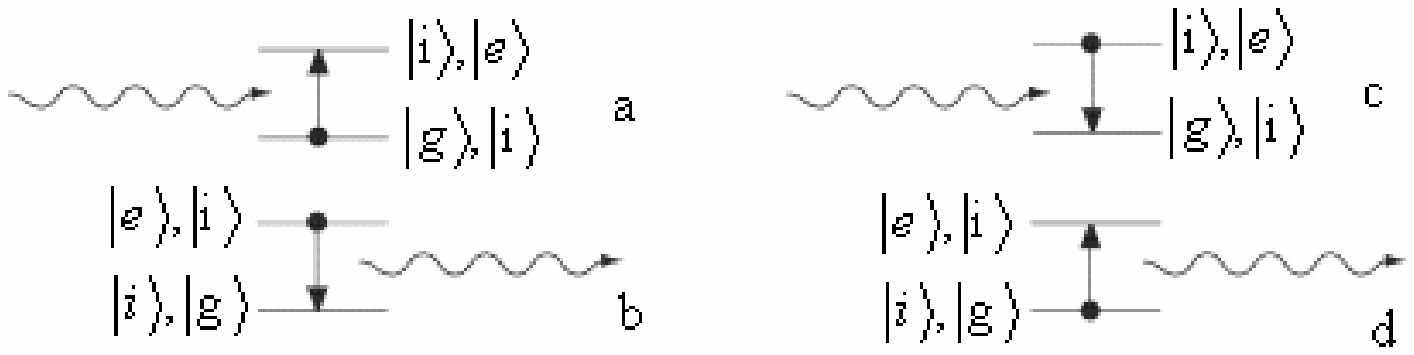}
\end{center}
\begin{center}
Figure 2.
\end{center}

On the contrary, the operators $\hat{\sigma}_{gi}\hat{a}^{+},$
$\hat{\sigma}_{ie}\hat{b}^{+}$ and $\hat{\sigma}_{ig}\hat{a}^{+},$
$\hat{\sigma}_{ei}\hat{b}$ either annihilate the excitation ($|i>$
or $|e>$) while creating the field excitation (Fig. 2b), or create
the excited atom (in $|i>$ or $|e>$) while annihilating the field
excitation (Fig. 2a).

Under the assumptions of real transitions, neglecting the
operators $\hat{\sigma}_{gi}\hat{a},$ $\hat{\sigma}_{ie}\hat{b},$
$\hat{\sigma}_{ig}\hat{a}^{+},$ $\hat{\sigma}_{ie}\hat{b}^{+},$ we
may reduce the Hamiltonian (\ref{ih5}) to the form
\begin{equation}\label{ih6}
\hat{H}_{\mathbf{r}\mathbf{E}}^L\cong\hat{H}_{int}^L=\hbar
g_a(\hat{\mathbf{R}})(\hat{\sigma}_{gi}\hat{a}^++\hat{\sigma}_{ig}\hat{a})+
\hbar g_b(\hat{\mathbf{R}})(\hat{\sigma}_{ie}\hat{b}^+
e^{i\delta\varphi}+\hat{\sigma}_{ei}\hat{b}e^{-i\delta\varphi}).
\end{equation}

\subsection{Simplification for $\Lambda$-configuration}
\textbf{Selection of real transitions.} For this configuration the
operators $\hat{\sigma}_{gi}\hat{a},$
$\hat{\sigma}_{ig}\hat{a}^{+},$ $\hat{\sigma}_{ie}\hat{b}^{+},$
$\hat{\sigma}_{ei}\hat{b}$ of the interaction Hamiltonian lead to
the violation of the energy conservation law (see Fig.1 and Fig.
3c,d). Thus, neglecting these terms in the interaction Hamiltonian
(\ref{ih5}) and keeping $\hat{\sigma}_{gi}\hat{a}^{+},$
$\hat{\sigma}_{ig}\hat{a},$ $\hat{\sigma}_{ie}\hat{b}$ and
$\hat{\sigma}_{ei}\hat{b}^{+},$ and (Fig. 3a,b), we may reduce the
Hamiltonian (\ref{ih5}) to the form
\begin{center}
\psfig{file=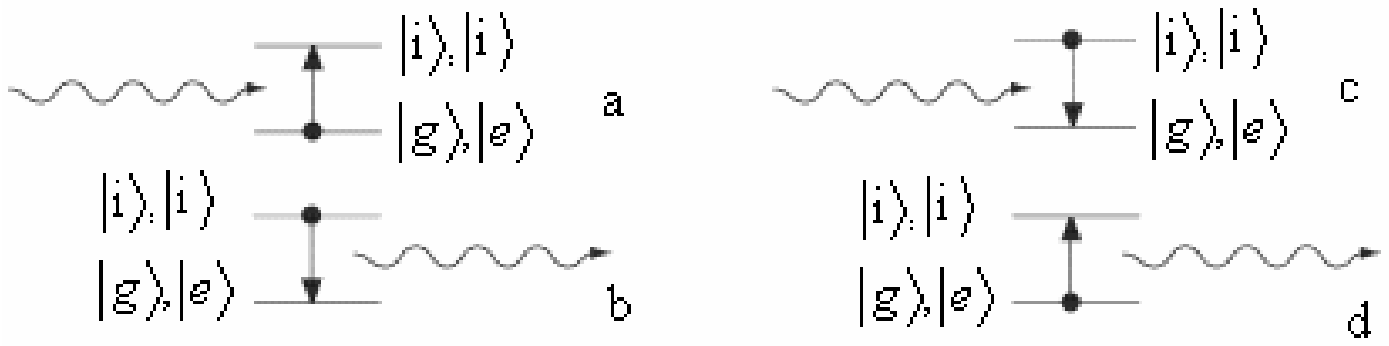}
\end{center}
\begin{center}
Figure 3.
\end{center}
\begin{equation}\label{ihl}
\hat{H}_{\mathbf{r}\mathbf{E}}^{\Lambda}\cong\hat{H}_{int}^L=\hbar
g_a(\hat{\mathbf{R}})(\hat{\sigma}_{gi}\hat{a}^++\hat{\sigma}_{ig}\hat{a})
-\hbar g_b(\hat{\mathbf{R}})(\hat{\sigma}_{ie}\hat{b}^+
e^{i\delta\varphi}+\hat{\sigma}_{ei}\hat{b}e^{-i\delta\varphi}).
\end{equation}

\subsection{Simplification for V-configuration}
\textbf{Selection of real transitions.} For this configuration the
operators $\hat{\sigma}_{gi}\hat{a}^{+},$
$\hat{\sigma}_{ig}\hat{a},$ $\hat{\sigma}_{ie}\hat{b},$ and
$\hat{\sigma}_{ei}\hat{b}^{+},$  of the interaction Hamiltonian
lead to the violation of the energy conservation law (see Fig.1
and Fig. 4c,d). Thus, neglecting these terms in the interaction
Hamiltonian (\ref{ih5}) and keeping $\hat{\sigma}_{gi}\hat{a},$
$\hat{\sigma}_{ig}\hat{a},$ $\hat{\sigma}_{ie}\hat{b},$ and
$\hat{\sigma}_{ei}\hat{b},$  and (Fig. 4a,b), we may reduce the
Hamiltonian (\ref{ih5}) to the form
\begin{center}
\psfig{file=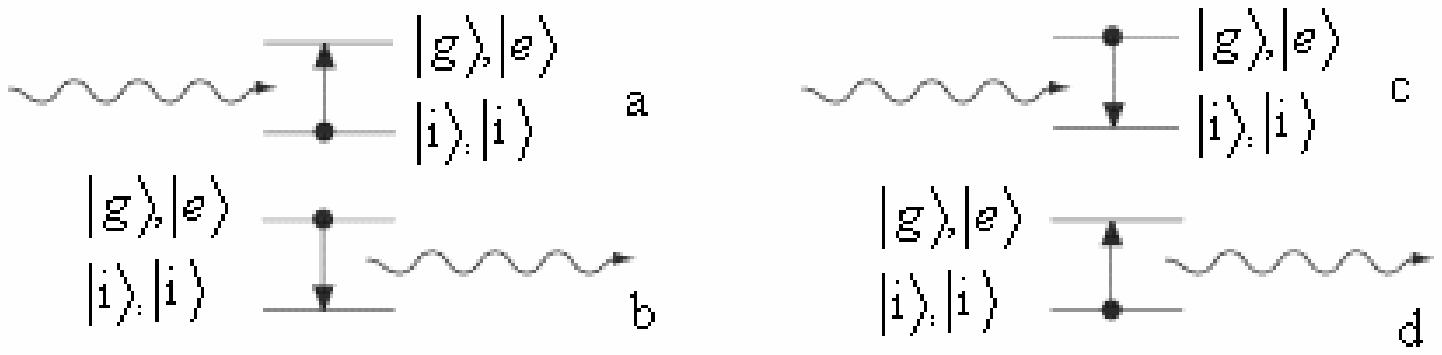}
\end{center}
\begin{center}
Figure 4.
\end{center}
\begin{equation}\label{ihv}
\hat{H}_{\mathbf{r}\mathbf{E}}^{\Lambda}\cong\hat{H}_{int}^V=-\hbar
g_a(\hat{\mathbf{R}})(\hat{\sigma}_{gi}\hat{a}^++\hat{\sigma}_{ig}\hat{a})+
\hbar g_b(\hat{\mathbf{R}})(\hat{\sigma}_{ie}\hat{b}^+
e^{i\delta\varphi}+\hat{\sigma}_{ei}\hat{b}e^{-i\delta\varphi}).
\end{equation}

\subsection{Interaction representation}
Further simplification of the atom-photon interaction hamiltonians
(\ref{ih4})-(\ref{ihv}) takes place in the interaction
representation, where we can use the averaging procedure by fast
oscillations. To accomplish this representation we should
represent the total Hamiltonian of the whole atom-field system
(see (\ref{total}))
$$
\hat{H}=\frac{\hat{\mathbf{P}}^2}{2M}+\hat{H}_{field}+\hat{H}_{atom}+\hat{H}_{\mathbf{r}\mathbf{E}}=
$$
$$
\frac{\hat{\mathbf{P}}^2}{2M}+\hat{H}_0+ \hbar
g_a(\hat{\mathbf{R}})(\hat{\sigma}_{ig}-\hat{\sigma}_{gi})(\hat{a}-\hat{a}^+)+
\hbar
g_b(\hat{\mathbf{R}})(\hat{\sigma}_{ei}e^{i\delta\varphi}-\hat{\sigma}_{ie}e^{-i\delta\varphi}(\hat{b}-\hat{b}^+).
$$
by the internal states of the atom and free-field states. Here
$$
\hat{H}_0=\hat{H}_{fild}+\hat{H}_{atom}=\hbar\Omega_a\hat{a}^+\hat{a}+\hbar\Omega_b\hat{b}^+\hat{b}+E_e\hat{\sigma}_{ee}
+E_i\hat{\sigma}_{ii}+E_g\hat{\sigma}_{gg},$$ or
\begin{equation}\label{ham0}
\hat{H}_0=\hbar\Omega_a\hat{a}^+\hat{a}+\hbar\Omega_b\hat{b}^+\hat{b}+
\hbar\left (
\begin{array}{cccc}
\omega_e& 0 & 0 \\
0 & \omega_i & 0 \\
0 & 0 & \omega_g
\end{array}
\right )
\end{equation}
denotes the sum of free-field $\hat{H}_{field}$  and free atom
$\hat{H}_{atom}$ Hamiltonians.

Now, defining the state vector in the interaction representation
$|\Phi^{(I)}(t)>$
 via \begin{equation}\label{s1}|\Phi(t)>\equiv
exp[-\frac{i}{\hbar}\hat{H}_0 t]|\Phi^{(I)}(t)>\end{equation} and
substituting (\ref{s1}), containing the state vector
$|\Phi^{(I)}(t)>$ in the interaction representation,  in the
Schr\"odinger equation
$$
i\hbar\frac{d|\Phi^{(I)}(t)>}{dt}=(\frac{\hat{\mathbf{P}}^2}{2M}+\hat{H}_0+\hat{H}_{\textbf{r}\textbf{E}})|\Phi(t)>.
$$
We receive
$$
i\hbar\frac{d|\Phi^{(I)}(t)>}{dt}=(\frac{\hat{\mathbf{P}}^2}{2M}+\hat{H}_{\textbf{r}\textbf{E}}^{I})|\Phi(t)>,
$$
where
\begin{equation}\label{hamI}
 \hat{H}_{\mathbf{r}\mathbf{E}}^{(I)}\equiv
\exp[\frac{i}{\hbar}\hat{H}_0t]\hat{H}_{\mathbf{r}\mathbf{E}}\exp[-\frac{i}{\hbar}\hat{H}_0t]
\end{equation}
is the interaction Hamiltonian in the interaction picture. Note,
that by reason of (vi) of section 2, we do not include the kinetic
energy operator (corresponding to the center of mass movement) in
above transformations. Now, let us obtain the exact expression for
the interaction Hamiltonian $\hat{H}_{\mathbf{r}\mathbf{E}}^{(I)}.$
Substituting $\hat{H}_0$ (\ref{ham0}) into the expression
(\ref{hamI}), under conditions, that the field and atom operators
commute with each other we obtain
\begin{equation}\label{hamI1}
\hat{H}_{\mathbf{r}\mathbf{E}}^{(I)}=\hbar
g_a(\mathbf{\hat{R}})e^{\frac{i}{\hbar}\hat{H}_{atom}t}(\hat{\sigma}_{ig}-\hat{\sigma}_{gi})e^{-\frac{i}{\hbar}\hat{H}_{atom}t}
e^{i\Omega_a\hat{a}^{+}\hat{a}t}(\hat{a}-\hat{a}^{+})e^{-i\Omega_a\hat{a}^{+}\hat{a}t}
\end{equation}
$$
+\hbar
g_a(\mathbf{\hat{R}})e^{\frac{i}{\hbar}\hat{H}_{atom}t}(\hat{\sigma}_{ei}e^{-i\delta\varphi}-\hat{\sigma}_{ie}e^{i\delta\varphi})
e^{-\frac{i}{\hbar}\hat{H}_{atom}t}
e^{i\Omega_b\hat{b}^{+}\hat{b}t}(\hat{b}-\hat{b}^{+})e^{-i\Omega_b\hat{b}^{+}\hat{b}t}.
$$
At first consider the part containing the atom operators. As the
matrix corresponding to $\hat{H}_{atom}$  is diagonal (see
(\ref{6}), (\ref{ham0})), then we obtain the following expression
\begin{equation}\label{expA}
exp(\frac{i}{\hbar}\hat{H}_{atom}t)=\left (
\begin{array}{cccc}
e^{i\omega_e t}& 0 & 0 \\
0 & e^{i\omega_i t} & 0 \\
0 & 0 & e^{i \omega_g t}
\end{array}
\right ).
\end{equation}
Utilizing (\ref{hamI1}) in (\ref{expA}) with the aid of (\ref{i1})
one can easily obtain that
$$
exp(\frac{i}{\hbar}\hat{H}_{atom}t)\hat{\sigma}_{ig}exp(-\frac{i}{\hbar}\hat{H}_{atom}t)\hat{\sigma}_{ig}=
$$
$$
=\hat{\sigma}_{ig}e^{i(\omega_i-\omega_g)t}
$$
In the same way, taking into account (\ref{i2}), (\ref{i3}) and
(\ref{i4})
$$
exp(\frac{i}{\hbar}\hat{H}_{atom}t)\hat{\sigma}_{gi}exp(-\frac{i}
{\hbar}\hat{H}_{atom}t)\hat{\sigma}_{ig}=
\hat{\sigma}_{gi}e^{-i(\omega_i-\omega_g)t},
$$
$$
exp(\frac{i}{\hbar}\hat{H}_{atom}t)\hat{\sigma}_{ei}e^{-i\delta\varphi}
exp(-\frac{i}{\hbar}\hat{H}_{atom}t)=
\hat{\sigma}_{ei}e^{i(\omega_e-\omega_i)t}e^{-i\delta\varphi},
$$
$$
exp(\frac{i}{\hbar}\hat{H}_{atom}t)\hat{\sigma}_{ie}e^{i\delta\varphi}
exp(-\frac{i}{\hbar}\hat{H}_{atom}t)\hat{\sigma}_{ig}=
\hat{\sigma}_{ig}e^{-i(\omega_e-\omega_i)t}e^{i\delta\varphi}.
$$
Thus, the atomic part of the interaction hamiltonian (\ref{hamI1})
is expressed in the following form
\begin{equation}\label{at1}
e^{\frac{i}{\hbar}\hat{H}_{atom}t}(\hat{\sigma}_{ig}-
\hat{\sigma}_{gi})e^{-\frac{i}{\hbar}\hat{H}_{atom}t}+
e^{\frac{i}{\hbar}\hat{H}_{atom}t}(\hat{\sigma}_{ei}e^{-i\delta\varphi}-\hat{\sigma}_{ie}
e^{i\delta\varphi}) e^{-\frac{i}{\hbar}\hat{H}_{atom}t}=
\end{equation}
$$
=\hat{\sigma}_{ig}e^{i(\omega_i-\omega_g)}t-\hat{\sigma}_{gi}
e^{-i(\omega_i-\omega_g)t}+\hat{\sigma}_{ei}e^{i(\omega_e-\omega_i)t}e^{-i\delta\varphi}
-\hat{\sigma}_{ei}e^{-i(\omega_e-\omega_i)t}e^{i\delta\varphi}.
$$
Now let us calculate the field operators. The time dependence of
the operators in the interaction representation is determined by
the Heisenberg equations of motion
\begin{equation}\label{heis1}
\frac{d}{dt}\hat{f}=\frac{i}{\hbar}[\hat{H}_{field},\hat{f}]\equiv
\frac{i}{\hbar}(\hat{H}_{filed}\hat{f}-\hat{f}\hat{H}_{filed}),
\end{equation}
where $\hat{f}$  stands for $\hat{a},$ $\hat{a}^{+},$ $\hat{b}$
and $\hat{b}$. Inserting $\hat{H}_{field}$ (\ref{qfh}) in
(\ref{heis1}) we obtain
\begin{equation}\label{heis2}
\frac{d}{dt}\hat{f}=\frac{i}{\hbar}\Omega(\hat{f}^{+}\hat{f}\hat{f}-\hat{f}\hat{f}^{+}\hat{f}).
\end{equation}
Because of the commutation relation (\ref{comm}) the equation
(\ref{heis2}) is transformed into
\begin{equation}\label{heis3}
\frac{d}{dt}\hat{f}=-i\Omega \hat{f}^{+}.
\end{equation}
In the same manner
\begin{equation}\label{heis4}
\frac{d}{dt}\hat{f}^+=i\Omega\hat{f}^+.
\end{equation}
So the time dependance of field operators is defined from
(\ref{heis3}) and (\ref{heis4}) and are equal
\begin{equation}\label{os}
\hat{a}(t)=\hat{a}(0)e^{-i\Omega_at}, \text{  }
\hat{a}(t)^{+}=\hat{a}^{+}(0)e^{i\Omega_at}, \text{  }
\hat{b}(t)=\hat{b}(0)e^{-i\Omega_bt}, \text{  }
\hat{b}(t)^{+}=\hat{b}^{+}(0)e^{i\Omega_bt},
\end{equation}
Thus, by substituting (\ref{at1}) and (\ref{os}) into the
expression (\ref{hamI1}) we obtain the atom-photon interactions
Hamiltonian in the interaction representation
$$
\hat{H}_{\mathbf{r}\mathbf{E}}^{(I)}=\hbar
g_a(\mathbf{\hat{R}})[\hat{\sigma}_{ig}e^{i(\omega_i-\omega_g)t}
-\hat{\sigma}_{gi}e^{-i(\omega_i-\omega_g)t}](\hat{a}e^{-i\Omega_at}-\hat{a}^{+}e^{\Omega_at})+
$$
$$
+\hbar
g_a(\mathbf{\hat{R}})[\hat{\sigma}_{ei}e^{i(\omega_e-\omega_i)t}e^{-i\delta\varphi}
-\hat{\sigma}_{ie}e^{-i(\omega_e-\omega_i)t}e^{i\delta\varphi}](\hat{b}e^{-i\Omega_bt}-\hat{b}^{+}e^{\Omega_bt}).
$$
Introduce the following notation
\begin{equation}\label{d}
\omega_i-\omega_j=\omega_{ij}.
\end{equation}
Taking into account (\ref{d}) the Hamiltonian   is given by the
following expression
$$
\hat{H}_{\mathbf{r}\mathbf{E}}^{(I)}=\hbar
g_a(\mathbf{\hat{R}})[\hat{\sigma}_{ig}e^{i\omega_{ig}t}
-\hat{\sigma}_{gi}e^{-i\omega_{ig}t}](\hat{a}e^{-i\Omega_at}-\hat{a}^{+}e^{i\Omega_at})+
$$
$$
+\hbar
g_b(\mathbf{\hat{R}})[\hat{\sigma}_{ei}e^{i\omega_{ei}t}e^{i\delta\varphi}
-\hat{\sigma}_{ie}e^{-i\omega_{ei}t}](\hat{b}e^{-i\Omega_bt}-\hat{b}^{+}e^{i\Omega_bt}),
$$
or
\begin{equation}\label{hamI2}
\hat{H}_{\mathbf{r}\mathbf{E}}^{(I)}=\hbar
g_a(\mathbf{\hat{R}})(\hat{\sigma}_{gi}\hat{a}^{+}e^{i(\Omega_a-\omega_{ig})t}
+\hat{\sigma}_{ig}\hat{a}e^{-i(\Omega_a-\omega_{ig})t}-\hat{\sigma}_{gi}\hat{a}e^{-i(\Omega_a+\omega_{ig})t}
-e^{-i\delta\varphi}\hat{\sigma}_{ig}\hat{a}^{+}e^{i(\Omega_a+\omega_{ig})t}+
\end{equation}
$$
+\hbar
g_b(\mathbf{\hat{R}})(\hat{\sigma}_{ie}\hat{b}^{+}e^{i(\Omega_b-\omega_{ei})t+i\delta\varphi}
+\hat{\sigma}_{ei}\hat{b}e^{-i(\Omega_b-\omega_{ei})t-i\delta\varphi}-
\hat{\sigma}_{ie}\hat{b}e^{-i(\Omega_b+\omega_{ei})t+i\delta\varphi}
-\hat{\sigma}_{ei}\hat{b}^{+}e^{i(\Omega_b+\omega_{ei})t-i\delta\varphi}.
$$
Further transformations require choosing particular transition
scheme (Fig. 1).

\

\textbf{L-configuration.} For this configuration (Fig.1)
\begin{equation}\label{d2}
\omega_g<\omega_i<\omega_e  \Rightarrow \omega_{ig}>0,
\omega_{ei}>0.
\end{equation}
After introducing notations $$\Delta_a\equiv
\Omega_a-\omega_{ig}$$ and
$$
\Delta_b\equiv \Omega_b-\omega_{ei}.
$$
from (\ref{hamI2}) and (\ref{d2}) we obtain the interaction
Hamiltonian for L-configuration
$\hat{H}_{\mathbf{r}\mathbf{E}}^{(I,L)}$
$$
\hat{H}_{\mathbf{r}\mathbf{E}}^{(I,L)}=\hbar
g_a(\mathbf{\hat{R}})(\hat{\sigma}_{gi}\hat{a}^{+}e^{i\Delta_a t}
+\hat{\sigma}_{ig}\hat{a}e^{-i\Delta_a
t}-\hat{\sigma}_{gi}\hat{a}e^{-i(\Omega_a+\omega_{ig})t}
-\hat{\sigma}_{ig}\hat{a}^{+}e^{i(\Omega_a+\omega_{ig})t})+
$$
$$
+\hbar
g_b(\mathbf{\hat{R}})(\hat{\sigma}_{ie}\hat{b}^{+}e^{i\Delta_bt+i\delta\varphi}
+\hat{\sigma}_{ei}\hat{b}e^{-i\Delta_b t -i\delta\varphi}
-\hat{\sigma}_{ie}\hat{b}e^{-i(\Omega_b+\omega_{ei})t+i\delta\varphi}
-\hat{\sigma}_{ei}\hat{b}^{+}e^{i(\Omega_b+\omega_{ei})t-i\delta\varphi}).
$$
It is easy to see that operators $\hat{\sigma}_{gi}\hat{a},$
$\hat{\sigma}_{ig}\hat{a}^{+},$ $\hat{\sigma}_{ie}\hat{b}$ and
$\hat{\sigma}_{ei}^{+}\hat{b}^{+}$ leading to the violation of
energy conservation law are multiplied by factors containing the
sums of field oscillating and atom transitions frequencies.
Contrary, operators $\hat{\sigma}_{gi}\hat{a}^{+},$
$\hat{\sigma}_{ig}\hat{a},$ $\hat{\sigma}_{ie}\hat{b}^{+}$ and
$\hat{\sigma}_{ei}\hat{b}$ are multiplied by factors containing the
differences (detuning) of aforementioned frequencies $\Delta\equiv
\Omega-\omega.$ So in the interaction Hamiltonian we have two groups
of terms. The terms of the first group oscillate approximately with
double optical frequencies and their contribution in the
Schr\"odinger equation is vanishingly small. The terms of another
group oscillate with detuning frequencies $\Delta$, and vary slowly.
So their contribution in the Schr\"odinger equation is significant.
These terms are $\hat{\sigma}_{gi}\hat{a}^{+},$
$\hat{\sigma}_{ig}\hat{a},$ $\hat{\sigma}_{ie}\hat{b}^{+}$ and
$\hat{\sigma}_{ei}\hat{b}.$ So utilize the averaging procedure -
neglecting the fast oscillating terms such as
$\hat{\sigma}_{gi}\hat{a},$ $\hat{\sigma}_{ig}\hat{a}^{+},$
$\hat{\sigma}_{ie}\hat{b}$ and $\hat{\sigma}_{ei}\hat{b}^{+}$  the
interaction Hamiltonian reduces to the following form
$$
\hat{H}_{\mathbf{r}\mathbf{E}}^{(I,L)}\cong\hat{H}_{int}^{L}=\hbar
g_a(\hat{\sigma}_{gi}\hat{a}^+e^{i\Delta_a t
}+\hat{\sigma}_{ig}\hat{a}e^{-i\Delta_a t })+\hbar
g_b(\hat{\sigma}_{ei}\hat{b}^+e^{i\Delta_b t+i\delta \varphi
}+\hat{\sigma}_{ei}\hat{b}e^{-i\Delta_b t- i\delta \varphi}).
$$
It is also possible to strongly validate given approach by the
mathematical averaging, contrary to heuristic averaging used above.
Also the high order contributions can be calculated in this case.

\

\textbf{$\Lambda$-configuration.} For this configuration (Fig.1)
$$
\omega_g<\omega_i<\omega_e  \Rightarrow \omega_{ig}>0,
\omega_{ei}<0.
$$
After introducing notations $$\Delta_a\equiv
\Omega_a-\omega_{ig}$$ and
$$
\Delta_b\equiv \Omega_b-\omega_{ie},
$$
we obtain the interaction Hamiltonian for $\Lambda$-configuration
$\hat{H}_{\mathbf{r}\mathbf{E}}^{(I,\Lambda)}$ in the interaction
representation
\begin{equation}\label{lam1}
\hat{H}_{\mathbf{r}\mathbf{E}}^{(I,\Lambda)}=\hbar
g_a(\mathbf{\hat{R}})(\hat{\sigma}_{gi}\hat{a}e^{i\Delta_at}
+\hat{\sigma}_{ig}\hat{a}e^{-i\Delta_at}-\hat{\sigma}_{gi}\hat{a}e^{-i(\Omega_a+\omega_{ig})t}
-\hat{\sigma}_{ig}\hat{a}^{+}e^{i(\Omega_a+\omega_{ig})t})+
\end{equation}
$$
+\hbar
g_b(\mathbf{\hat{R}})(\hat{\sigma}_{ie}\hat{b}^{+}e^{i(\Omega_b+\omega_{ie})t+i\delta\varphi}
+\hat{\sigma}_{ei}\hat{b}e^{-i\Delta_b t +i\delta\varphi}
-\hat{\sigma}_{ei}\hat{b}e^{i\Delta_bt+i\delta\varphi}
-\hat{\sigma}_{ei}\hat{b}^{+}e^{i\Delta_bt-i\delta\varphi}).
$$
Averaging (\ref{lam1}) by fast oscillations leads to
$$
\hat{H}_{\mathbf{r}\mathbf{E}}^{(I,\Lambda)}\cong\hat{H}_{int}^{\Lambda}=\hbar
g_a(\hat{\sigma}_{gi}^-\hat{a}^+e^{i\Delta_a t
}+\hat{\sigma}_{gi}\hat{a}e^{-i\Delta_a t })+\hbar
g_b(\hat{\sigma}_{ie}^-\hat{b}^+e^{-i\Delta_b t+i\delta \varphi
}+\hat{\sigma}_{ei}\hat{b}e^{i\Delta_b t- i\delta \varphi}).
$$

\

\textbf{$V$-configuration.} For this configuration (Fig.1)
$$
\omega_g<\omega_i<\omega_e  \Rightarrow \omega_{ig}<0,
\omega_{ei}>0.
$$
After introducing notations $$\Delta_a\equiv
\Omega_a-\omega_{gi}$$ and
$$
\Delta_b\equiv \Omega_b-\omega_{ei},
$$
we obtain the interaction Hamiltonian for $V$-configuration
$\hat{H}_{\mathbf{r}\mathbf{E}}^{(I,V)}$ in the interaction
representation
\begin{equation}\label{hamIV}
\hat{H}_{\mathbf{r}\mathbf{E}}^{(I,\Lambda)}=\hbar
g_a(\mathbf{\hat{R}})(\hat{\sigma}_{gi}\hat{a}^{+}e^{i(\Omega_a+\omega_{gi})
t} +\hat{\sigma}_{ig}\hat{a}e^{-i(\Omega_a+\omega_{gi})
t}-\hat{\sigma}_{gi}\hat{a}e^{-i\Delta_at}
-\hat{\sigma}_{ig}\hat{a}^{+}e^{i\Delta_at}+
\end{equation}
$$
+\hbar
g_b(\mathbf{\hat{R}})(\hat{\sigma}_{ie}\hat{b}^{+}e^{i\Delta_bt+i\delta\varphi}
+\hat{\sigma}_{ei}\hat{b}e^{-i\Delta_b t -i\delta\varphi}
-\hat{\sigma}_{ie}\hat{b}e^{-i(\Omega_b+\omega_{ei})t+i\delta\varphi}
-\hat{\sigma}_{ei}\hat{b}^{+}e^{-i(\Omega_b+\omega_{ei})t-i\delta\varphi}).
$$
Neglecting in (\ref{hamIV}) fast oscillating terms we have
$$
\hat{H}_{\mathbf{r}\mathbf{E}}^{(I,V)}\cong\hat{H}_{int}^{V}=-\hbar
g_a(\hat{\sigma}_{gi}\hat{a}e^{-i\Delta_a t
}+\hat{\sigma}_{gi}\hat{a}^{+}e^{i\Delta_a t })+\hbar
g_b(\hat{\sigma}_{ie}\hat{b}^+e^{i\Delta_b t+i\delta \varphi
}+\hat{\sigma}_{ei}\hat{b}e^{-i\Delta_b t- i\delta \varphi}).
$$

Therefore, in this section we have shown that in the RWA
three-level atom interacting with two modes of standing wave
resonator, in the interaction representation is described by the
following Hamiltonians

L-configuration
\begin{equation}\label{L}
\hat{H}_{\mathbf{r}\mathbf{E}}^{(I,L)}\cong\hat{H}_{int}^L=\hbar
g_a(\hat{\sigma}_{gi}\hat{a}^+e^{-i\Delta_a t
}+\hat{\sigma}_{ig}\hat{a}e^{-i\Delta_a t })+\hbar
g_b(\hat{\sigma}_{ie}\hat{b}^+e^{i\Delta_b t+i\delta \varphi
}+\hat{\sigma}_{ei}\hat{b}e^{-i\Delta_b t- i\delta \varphi}).
\end{equation}

$\Lambda$-configuration
\begin{equation}\label{La}
\hat{H}_{\mathbf{r}\mathbf{E}}^{(I,\Lambda)}\cong\hat{H}_{int}^{\Lambda}=\hbar
g_a(\hat{\sigma}_{gi}\hat{a}^+e^{i\Delta_a t
}+\hat{\sigma}_{ig}\hat{a}e^{-i\Delta_a t })+\hbar
g_b(\hat{\sigma}_{ie}\hat{b}e^{-i\Delta_b t+i\delta \varphi
}+\hat{\sigma}_{ei}\hat{b}e^{i\Delta_b^+ t- i\delta \varphi}).
\end{equation}

V-configuration
\begin{equation}\label{V}
\hat{H}_{\mathbf{r}\mathbf{E}}^{(I,V)}\cong\hat{H}_{int}^{V}=-\hbar
g_a(\hat{\sigma}_{gi}\hat{a}e^{-i\Delta_a t
}+\hat{\sigma}_{ig}\hat{a}^+e^{i\Delta_a t })+\hbar
g_b(\hat{\sigma}_{ie}\hat{b}^+e^{i\Delta_b t+i\delta \varphi
}+\hat{\sigma}_{ei}\hat{b}e^{-i\Delta_b t- i\delta \varphi}).
\end{equation}
Farther we will investigate three-level model concentrated on the
internal degrees of freedom and neglecting center of mass
movement.

\section{The Quantum Dynamics of a Three-Level Atom}
In this section we consider the internal quantum dynamics of a
three-level atom in the fields of two modes of standing wave
resonator. This model is more complicated and less studied than the
well-known two-level system (see \cite{shleikh}, \cite{allen} and
references cited there). So, consider the Schr\"odinger equation
\begin{equation}\label{sh1}
i\hbar\frac{d|\Psi(t)>}{dt}=\hat{H}_{int}|\Psi(t)>
\end{equation}
for the state vector $|\Psi>$  in the interaction representation,
where the interaction Hamiltonian $\hat{H}_{int}$   due to
(\ref{L})-(\ref{V}) contains the following terms
\begin{equation}\label{L1}
\hat{H}_{int}=\hat{H}_{int}^L=\hbar
g_a(\hat{\sigma}_{gi}^-\hat{a}^+e^{i\Delta_a t
}+\hat{\sigma}_{gi}\hat{a}e^{i\Delta_a t })+\hbar
g_b(\hat{\sigma}_{ie}^-\hat{b}^+e^{i\Delta_b t+i\delta \varphi
}+\hat{\sigma}_{ei}\hat{b}e^{-i\Delta_b t- i\delta \varphi}),
\end{equation}
\begin{equation}\label{La1}
\hat{H}_{int}=\hat{H}_{int}^{\Lambda}=\hbar
g_a(\hat{\sigma}_{gi}^-\hat{a}^+e^{i\Delta_a t
}+\hat{\sigma}_{gi}\hat{a}e^{-i\Delta_a t })+\hbar
g_b(\hat{\sigma}_{ie}^-\hat{b}^+e^{-i\Delta_b t+i\delta \varphi
}+\hat{\sigma}_{ei}\hat{b}e^{i\Delta_b t- i\delta \varphi}),
\end{equation}
\begin{equation}\label{V1}
\hat{H}_{int}=\hat{H}_{int}^{V}=-\hbar
g_a(\hat{\sigma}_{gi}^-\hat{a}^+e^{-i\Delta_a t
}+\hat{\sigma}_{gi}\hat{a}e^{i\Delta_a t })+\hbar
g_b(\hat{\sigma}_{ie}^-\hat{b}^+e^{i\Delta_b t+i\delta \varphi
}+\hat{\sigma}_{ei}\hat{b}e^{-i\Delta_b t- i\delta \varphi}).
\end{equation}
We will study the internal dynamics of the three-level system when
the interaction Hamiltonian becomes time independent. It is the most
convenient way to obtain the time evolution operator.

\subsection{Exact resonance in three-level atom.}
In the case of exact resonance $\Delta=0$  the interaction
Hamiltonians (\ref{L1})-(\ref{V1}) transform into the following ones
\begin{equation}\label{f1}
\hat{H}_{int}=\hat{H}_{int}^L=\hbar
g_a(\hat{\sigma}_{gi}\hat{a}^++\hat{\sigma}_{ig}\hat{a})+\hbar
g_b(\hat{\sigma}_{ie}\hat{b}^+e^{i\delta \varphi
}+\hat{\sigma}_{ei}\hat{b}e^{-i\delta \varphi}),
\end{equation}
\begin{equation}\label{f2}
\hat{H}_{int}=\hat{H}_{int}^{\Lambda}=\hbar
g_a(\hat{\sigma}_{gi}\hat{a}^++\hat{\sigma}_{gi}\hat{a})-\hbar
g_b(\hat{\sigma}_{ie}\hat{b}e^{i\delta \varphi
}+\hat{\sigma}_{ei}\hat{b}^+e^{- i\delta \varphi}),
\end{equation}
\begin{equation}\label{f3}
\hat{H}_{int}=\hat{H}_{int}^{V}=-\hbar
g_a(\hat{\sigma}_{gi}\hat{a}^++\hat{\sigma}_{ig}\hat{a}^+)+\hbar
g_b(\hat{\sigma}_{ie}\hat{b}^+e^{i\delta \varphi
}+\hat{\sigma}_{ei}\hat{b}e^{- i\delta \varphi}).
\end{equation}
Now they do not contain the explicit time dependence and the
Schr\"odinger equation (\ref{sh1}) for (\ref{f1})-(\ref{f3}) can be
integrated formally to give
$$
|\Psi(t)>=exp(-\frac{i}{\hbar}\hat{H}_{int}t)|\Psi(0)>,
$$
where  the unitary time evolution operator is defined as
\begin{equation}\label{te}
\hat{U}(t,t_0=0)\equiv exp(-\frac{i}{\hbar}\hat{H}_{int}t)
\end{equation}
Substituting in (\ref{te}) the Hamiltonians (\ref{f1})-(\ref{f3}),
we obtain the time evolution operators for $L$-, $\Lambda$- and
$V$-configurations of a three-level atom
\begin{equation}\label{t1}
\hat{U}^L(t,t_0=0)=\exp
[-ig_at(\hat{\sigma}_{gi}\hat{a}^++\hat{\sigma}_{ig}\hat{a})-i\hbar
g_bt(\hat{\sigma}_{ie}\hat{b}^+e^{i\delta \varphi
}+\hat{\sigma}_{ei}\hat{b}e^{-i\delta \varphi})],
\end{equation}
\begin{equation}\label{t2}
\hat{U}^{\Lambda}(t,t_0=0)=\exp [-i
g_at(\hat{\sigma}_{gi}\hat{a}^++\hat{\sigma}_{ig}\hat{a})-i
g_bt(\hat{\sigma}_{ie}\hat{b}^+e^{i\delta \varphi
}+\hat{\sigma}_{ei}\hat{b}e^{- i\delta \varphi})],
\end{equation}
\begin{equation}\label{t3}
\hat{U}^{V}(t,t_0=0)=\exp[i
g_at(\hat{\sigma}_{gi}\hat{a}^++\hat{\sigma}_{gi}\hat{a})-i\hbar
g_bt(\hat{\sigma}_{ie}\hat{b}^+e^{i\delta \varphi
}+\hat{\sigma}_{ei}\hat{b}e^{- i\delta \varphi})].
\end{equation}
The initial state vector
\begin{equation}\label{ten}
|\Psi(0)\equiv |\Psi_{field}>\otimes |\Psi_{atom}>=
\end{equation}
$$
\sum_{n_a=0}^{\infty}\sum_{n_b=0}^{\infty}w_{n_a}w_{n_b}[c_e|e>+c_i|i>+c_g|g>]|n>=\sum_{n_a=0}^{\infty}\sum_{n_b=0}^{\infty}w_{n_a}w_{n_b}
\left (
\begin{array}{cccc}
c_e \\
c_i \\
c_g
\end{array}
\right )|n_a>|n_b>,
$$
represents the direct product of the field state vector
\begin{equation}\label{ten2}
|\Psi_{field}>\equiv\sum_{n_a=0}^{\infty}\sum_{n_b=0}^{\infty}w_{n_a}w_{n_b}|n_a>|n_b>
\end{equation}
and the atom state vector
\begin{equation}\label{89}
|\Phi_{atom}>\equiv[c_e|e>+c_i|i>+c_g|g>].
\end{equation}

It is a convenient form for utilizing the unitary transformations of
initial atom-field state vector. So, the dynamics of a three-level
atom, in the case of exact resonance, is defined by the
(\ref{t1})-(\ref{t3}) time evolution operators. In the next section
we calculate these operators and utilize the unitary transformations
important for applications of quantum computation.

\section{Calculation of the Time Evolution Operator and Unitary Transformations}

\subsection{Calculation of the time evolution operator in terms of the exponential matrix}
Based on relations (\ref{i1})-(\ref{i3}) let us rewrite the time
evolution operators (\ref{t1})-(\ref{t3}) in the matrix form
\begin{equation}\label{u1}
\hat{U}^{(L)}(t,t_0=0)=e^{-i\left (
\begin{array}{cccc}
0& e^{-i\delta \varphi} g_bt\hat{b} & 0 \\
e^{i\delta \varphi} g_bt\hat{b}^{+} & 0 &  g_at\hat{a} \\
0 & g_at\hat{a}^+ & 0
\end{array}
\right )}
\end{equation}
\begin{equation}\label{u2}
\hat{U}^{(\Lambda)}(t,t_0=0)=e^{-i\left (
\begin{array}{cccc}
0& -e^{-i\delta \varphi} g_bt\hat{b}^+ & 0 \\
-e^{i\delta \varphi} g_bt\hat{b} & 0 &  g_at\hat{a} \\
0 & g_at\hat{a}^{+} & 0
\end{array}
\right )},
\end{equation}
\begin{equation}\label{u3}
\hat{U}^{(V)}(t,t_0=0)=e^{-i\left (
\begin{array}{cccc}
0& e^{-i\delta \varphi} g_bt\hat{b} & 0 \\
e^{i\delta \varphi} g_bt\hat{b}^{+} & 0 & -g_at\hat{a}^+ \\
0 & -g_at\hat{a} & 0
\end{array}
\right )},
\end{equation}
or in the general form as
$$
\hat{U}(t,t_0=0)=e^{-i\left (
\begin{array}{cccc}
0& \hat{B} & 0 \\
\hat{B}^{+} & 0 &  \hat{A} \\
0 & \hat{A}^{+} & 0
\end{array}
\right )},
$$
where
$$
\begin{array}{ccccccc}
&\texttt{configurations:} & L& \Lambda & V \\
& & &&\\
\hat{A}&=&g_at\hat{a}&  g_at\hat{a}& -g_at\hat{a}^{+}\\
\hat{A}^{+}&=&g_at\hat{a}^{+}&  g_at\hat{a}^{+}& -g_at\hat{a}\\
\hat{B}&=&e^{-i\delta\varphi}g_bt\hat{b}&  -e^{i\delta\varphi}g_bt\hat{b}^{+}& e^{-i\delta\varphi}g_bt\hat{b}\\
\hat{B}^{+}&=&e^{i\delta\varphi}g_bt\hat{b}^+&  -e^{-i\delta\varphi}g_bt\hat{b}^{+}& e^{i\delta\varphi}g_bt\hat{b}^+\\
\end{array}
$$
corresponding operators from (\ref{u1})-(\ref{u3}), including the
corresponding scalar factors for convenience.

\subsection{Semiclassical approach}
At the given stage we substantially simplify the situation by means
of simplifying the field operators. Right hand sides of equations
(\ref{os}) include only $\hat{f}\exp{i\Omega t}$ type terms, which
describe contribution due to free evolution of the field. On the
other hand it means that the applied fields are sufficiently intense
compared with what can be contributed by the atomic transitions. The
field operators at time $t$ may then be replaced by their averages
in the initial state \cite{puri}. We assume the fields to be in the
coherent states (the field of laser radiation) $|\{\alpha_n \}>$ and
$|\{\beta_n \}>$ ($|\alpha|,|\beta|>>1$), replace the operators
$\hat{a}$ $(t_0=0)$ and $b$ $(t_0=0)$ by $\alpha$ and $\beta$ in
(\ref{u1})-(\ref{u3}) and receive the following

$L$-configuration
$$
\hat{U}^{(L)}(t,t_0=0)=e^{-i\left (
\begin{array}{cccc}
0& e^{-i\delta \varphi} g_{\beta}t\beta & 0 \\
e^{i\delta \varphi} g_{\beta}t\beta^{*} & 0 &  g_at\alpha \\
0 & g_at\alpha^* & 0
\end{array}
\right )},
$$
$\Lambda$-configuration
$$
\hat{U}^{(\Lambda)}(t,t_0=0)=e^{-i\left (
\begin{array}{cccc}
0& -e^{-i\delta \varphi} g_{\beta}t\beta^{*} & 0 \\
-e^{i\delta \varphi} g_{\beta}t\beta & 0 &  g_{\alpha}t\alpha \\
0 & g_{\alpha}t\alpha^{*} & 0
\end{array}
\right )},
$$
$V$-configuration
$$
\hat{U}^{(V)}(t,t_0=0)=e^{-i\left (
\begin{array}{cccc}
0& e^{-i\delta \varphi} g_{\beta}t\beta & 0 \\
e^{i\delta \varphi} g_{\beta}t\beta^{*} & 0 & -g_{\alpha}t\alpha^{*} \\
0 & -g_{\alpha}t\alpha & 0
\end{array}
\right )}.
$$
In general we have
 \begin{equation}\label{gen}
 \hat{U}(t,t_0=0)=e^{-i\left (
\begin{array}{cccc}
0& B & 0 \\
B^{*} & 0 &  A \\
0 & A^{*} & 0
\end{array}
\right )},
\end{equation}
where
\begin{equation}\label{tab}
\begin{array}{ccccccc}
&\texttt{configurations:} & L& \Lambda & V \\
& & &&\\
A&=&g_{\alpha}t\alpha&  g_{\alpha}t\alpha& -g_{\alpha}t\alpha^{*}\\
A^{*}&=&g_{\alpha}ta^{*}&  g_{\alpha}ta^{*}& -g_{\alpha}t\alpha\\
B&=&e^{-i\delta\varphi}g_{\beta}t\beta&  -e^{-i\delta\varphi}g_{\beta}t\beta^{*}& e^{-i\delta\varphi}g_{\beta}t\beta\\
B^{*}&=&e^{i\delta\varphi}g_{\beta}t\beta^*&  -e^{-i\delta\varphi}g_{\beta} t\beta^{*}& e^{i\delta\varphi}g_{\beta}t\beta^{*}\\
\end{array}
\end{equation}

The exponential matrix (\ref{gen}) can be calculated to give
\begin{equation}\label{5u}
\hat{U}(t,t_0=0)={\left (
\begin{array}{cccc}
\frac{AA^{*}+BB^{*}cos(\sqrt{BB^{*}+AA^{*}})}{BB^{*}+AA^{*}}& -\frac{iBsin(\sqrt{BB^{*}+AA^{*}})}{\sqrt{BB^{*}+AA^{*}}} & -\frac{BA(cos(\sqrt{BB^{*}+AA^{*}})-1)}{BB^{*}+AA^{*}} \\
-\frac{iB^*sin(\sqrt{BB^{*}+AA^{*}})}{\sqrt{BB^{*}+AA^{*}}} &cos(\sqrt{BB^{*}+AA^{*}}) & -\frac{iAsin(\sqrt{BB^{*}+AA^{*}})}{\sqrt{BB^{*}+AA^{*}}} \\
\frac{A^{*}B^{*}(cos(\sqrt{BB^{*}+AA^{*}})-1)}{BB^{*}+AA^{*}} &
-\frac{iA^{*}sin(\sqrt{BB^{*}+AA^{*}})}{\sqrt{BB^{*}+AA^{*}}}&\frac{BB^{*}+AA^{*}cos(\sqrt{BB^{*}+AA^{*}})}{BB^{*}+AA^{*}}
\end{array}
\right )},
\end{equation}

\subsection{Unitary transformations}
So, we found the time evolution operator for the three-level atom
interacting with two modes of field in the coherent states. As the
initial state vector is represented for the field in the arbitrary
state $
|\Psi_{field}>\equiv\sum_{n_a=0}^{\infty}\sum_{n_b=0}^{\infty}w_{n_a}w_{n_b}|n_a>|n_b>,
$ than it is essential to rewrite this state vector for two modes in
coherent state. The probability  that there are $n_a$ $(n_b)$
photons at the initial coherent state $|\alpha>$ $(|\beta>)$ is
given by the Poisson distribution \cite{Scully}, i.~e.
\begin{equation}\label{poi}
w_{n_{a}}=\frac{<n_a>^{n_a}e^{-<n_a>}}{n_a!},
w_{n_{b}}=\frac{<n_b>^{n_b}e^{-<n_b>}}{n_b!},
\end{equation}
where  $<n>=|\alpha|^2$ $(|\beta|^2)$. Due to (\ref{poi}) we can now
write the initial state of field  (\ref{ten2}) and atom-field system
(\ref{ten}) in more suitable form
\begin{equation}\label{5f}
|\Psi_{field}>\equiv\sum_{n_a=0}^{\infty}\sum_{n_b=0}^{\infty}\frac{<n_a>^{n_a}e^{-<n_a>}}{n_a!}
\frac{<n_b>^{n_b}e^{-<n_b>}}{n_b!}|n_a>|n_b>=|\alpha>|\beta>,
\end{equation}
\begin{equation}\label{5p}
|\Psi(0)>=|\Psi_{field}>\otimes|\Psi_{atom}>=
\end{equation}
$$
=\sum_{n_a=0}^{\infty}\sum_{n_b=0}^{\infty}\frac{<n_a>^{n_a}e^{-<n_a>}}{n_a!}
\frac{<n_b>^{n_b}e^{-<n_b>}}{n_b!}[c_e|e>+c_i|i>+c_g|g>]|n_a>|n_b>=
$$
$$
=\sum_{n_a=0}^{\infty}\sum_{n_b=0}^{\infty}\frac{<n_a>^{n_a}e^{-<n_a>}}{n_a!}
\frac{<n_b>^{n_b}e^{-<n_b>}}{n_b!}\left (
\begin{array}{cccc}
c_e \\
c_i\\
c_g
\end{array}
\right)|a_a>|n_b>=
$$
$$
=[c_e|e>+c_i|i>+c_g|g>]|\alpha>|\beta>=\left (
\begin{array}{cccc}
c_e \\
c_i\\
c_g
\end{array}
\right)|\alpha>|\beta>.
$$
Now applying (\ref{5u}) to (\ref{5p}) we obtain the total state
vector $|\Psi(t)>$ of a three-level atom in the coherent field of
two modes for arbitrary time $t>t_0=0$
\begin{equation}\label{5.3}
|\Psi(t)>=\hat{U}(t,t_0=0)[c_e|e>+c_i|i>+c_g|g>]|\alpha>|\beta>=\hat{U}(t,t_0=0)\left
(
\begin{array}{cccc}
c_e \\
c_i\\
c_g
\end{array}
\right)|\alpha>|\beta>.
\end{equation}

\subsection{Unitary transformations for L-configuration} For
L-configuration (\ref{89}), (\ref{5.3})  gives
$$
|\Psi(t)>=\hat{U}^{(L)}(t,t_0=0)[c_e|e>+c_i|i>+c_g|g>]|\alpha>|\beta>=\hat{U}^{(L)}(t,t_0=0)\left
(
\begin{array}{cccc}
c_e \\
c_i\\
c_g
\end{array}
\right)|\alpha>|\beta>.
$$
and due to (\ref{5u}) and (\ref{tab}) we obtain
$$
\hat{U}^{(L)}(t,t_0=0)=\frac{1}{g_a^2|\alpha|^2+g_b^2|\beta|^2}\times
\left \{ \left ( \begin{array}{cccc}
g_a^2|\alpha|^2 & 0& -e^{-i\delta\varphi}g_ag_b\beta\alpha\\
0&0&0\\
-e^{i\delta\varphi}g_ag_b\beta^{*}\alpha^{*}&0&g_b^2|\beta|^2
\end{array}
\right ) +\right.
$$
$$
+cos(t\sqrt{g_a^2|\alpha|^2+g_b^2|\beta|^2}) \left (
\begin{array}{cccc}
g_b^2|\beta|^2 & 0& e^{-i\delta\varphi}g_ag_b\beta\alpha\\
0&g_a^2\alpha^2+g_b^2\beta^2&0\\
e^{i\delta\varphi}g_ag_b\beta^{*}\alpha^{*}&0&g_a^2|\alpha|^2
\end{array}
\right )-
$$
$$
\left. -sin(t\sqrt{g_a^2|\alpha|^2+g_b^2|\beta|^2}) \left (
\begin{array}{cccc}
0 & e^{-i\delta\varphi}g_b\beta& 0\\
e^{i\delta\varphi}g_b\beta^{*}&0&g_a\alpha\\
0&g_a\alpha^{*}&0
\end{array}
 \right )\right \}.
$$
Thus, from the time evolution operator one can extract three formal
parts: time independent, time dependent as $\cos(Gt)$ and time
dependent as $\sin(Gt).$  For the unitary operator
$\hat{U}^{(L)}(t,t_0=0)$ an initial moment $t=0$  we obtain
\begin{equation}\label{pr1}
\hat{U^{(L)}}(t=0)= \left (
\begin{array}{cccc}
1 & 0& 0\\
0&1&0\\
0&0&1
\end{array}
 \right )
\end{equation}
and
\begin{equation}\label{pr2}
\Psi(t=0)= \left (
\begin{array}{cccc}
c_e(0)|\alpha>|\beta>\\
c_i(0)|\alpha>|\beta>\\
c_g(0)|\alpha>|\beta>
\end{array}
 \right )
\end{equation}
or
\begin{equation}\label{pr3}
\Psi(t=0)=c_e(0)|e,\alpha,\beta>+c_i(0)|i,\alpha,\beta>+c_g|g,\alpha,\beta>,
\end{equation}
where the following notations are introduced
$|e>|\alpha>|\beta>\equiv |e,\alpha,\beta>,$
$|i>|\alpha>|\beta>\equiv |i,\alpha,\beta>$ and
$|g>|\alpha>|\beta>\equiv |g,\alpha,\beta>.$ (\ref{pr1})-(\ref{pr3})
confirm the main feature of the interaction representation, that in
the absence of interaction the state vectors do not change
\cite{Cohen-Tannoudji}.

\section{Probability Amplitude Method}
In this section, we present a different but equivalent method to
solve for the evolution of the atom-field system described in
interaction picture
\begin{equation}\label{sh7.1}
i\hbar\frac{d|\Psi(t)>}{dt}=\hat{H}_{int}|\Psi(t)>
\end{equation}
and based on the solutions of the probability amplitudes. The
atom-field system is described by the interaction Hamiltonian
(\ref{hamI2})
\begin{equation}\label{7.2}
 \hat{H}_{\mathbf{r}\mathbf{E}}^{(I)}=\hbar
g_a(\hat{\sigma}_{gi}\hat{a}^{+}e^{i(\Omega_a-\omega_{ig})t}
+\hat{\sigma}_{ig}\hat{a}e^{-i(\Omega_a-\omega_{ig})t}-\hat{\sigma}_{gi}
\hat{a}e^{-i(\Omega_a+\omega_{ig})t}
+\hat{\sigma}_{ig}\hat{a}^{+}e^{i(\Omega_a+\omega_{ig})t}+
\end{equation}
$$
+\hbar
g_b(\hat{\sigma}_{ie}\hat{b}^{+}e^{i(\Omega_b-\omega_{ei})t+i\delta\varphi}
+\hat{\sigma}_{ei}\hat{b}e^{-i(\Omega_b-\omega_{ei})t-i\delta\varphi}-\hat{\sigma}_{ie}\hat{b}e^{-i(\Omega_b+\omega_{ei})t+i\delta\varphi}
-\hat{\sigma}_{ei}\hat{b}^{+}e^{i(\Omega_b+\omega_{ei})t-i\delta\varphi}.
$$
where the RWA should be utilized. Further calculations are based
on amplitude method of QED \cite{Scully} and are very similar for
different configurations. So we perform the calculations only for
L-configuration and then present the final result for all of them.

\subsection{L-Configuration}
In the RWA, the Hamiltonian (\ref{7.2}), due to configuration (see
Fig. 1), transforms into the ((\ref{L1})-(\ref{V1})) interaction
operators. Especially for L-configuration
\begin{equation}\label{7L}
\hat{H}_{int}=\hat{H}_{int}^L=\hbar
g_a(\hat{\sigma}_{gi}\hat{a}^+e^{i\Delta_a t
}+\hat{\sigma}_{ig}\hat{a}e^{-i\Delta_a t })+\hbar
g_b(\hat{\sigma}_{ie}\hat{b}^+e^{i\Delta_b t+i\delta \varphi
}+\hat{\sigma}_{ei}\hat{b}e^{-i\Delta_b t- i\delta \varphi}).
\end{equation}
 In
the exact resonances case ($\Delta=0$ ) (\ref{7L}) transforms into
\begin{equation}\label{7ll}
\hat{H}_{int}^L=\hbar
g_a(\hat{\sigma}_{gi}\hat{a}^++\hat{\sigma}_{ig}\hat{a})+\hbar
g_b(\hat{\sigma}_{ie}\hat{b}^+e^{i\delta\varphi}+\hat{\sigma}_{ei}\hat{b}e^{-
i\delta \varphi}). \end{equation}

Now acting on the wave vector $|e,n_a,n_b>$  by the operator
(\ref{7ll}) step by step we obtain
\begin{equation}\label{l11}
\hat{H}_{int}^L|e,n_a,n_b>=\hbar
g_be^{i\delta\varphi}\sqrt{n_b+1}|e,n_a,n_b>,
\end{equation}
\begin{equation}\label{l12}
\hat{H}_{int}^L|e,n_a,n_b+1>=\hbar
g_a\sqrt{n_a+1}|g,n_a+1,n_b+1>+\hbar
g_be^{-i\delta\varphi}\sqrt{n_b+1}|e,n_a,n_b>,
\end{equation}
\begin{equation}\label{l13}
\hat{H}_{int}^L|g,n_a+1,n_b+1>=\hbar g_a\sqrt{n_a+1}|i,n_a,n_b+1>,
\end{equation}
where $n_a$ and $n_b$  represent the number of $\hbar\Omega_a$ and
$\hbar\Omega_b$ photons. From (\ref{l13}) it is also easy to see
that for $n_a=0$ and for arbitrary $n_b$ (let us $n_b+1=m$)
\begin{equation}\label{l14}
\hat{H}_{int}^L|g,0,m>=0.
\end{equation}
Now with the aid of obtained amplitudes (\ref{l11})-(\ref{l14}) one
proceeds to solve the equation of motion (\ref{sh7.1}) for state
vector of total system $|\Psi(t)>$ , i.~e.,
\begin{equation}\label{st1}
|\Psi(t)>=\sum_{n_a=0,n_b=0}^{\infty}\Psi_{e,n_a,n_b}(t)|e,n_a,n_b>+\Psi_{i,n_a,n_b+1}(t)|i,n_a,n_b+1>+
\end{equation}
$$
+\Psi_{g,n_a+1,n_b+1}(t)|g,n_a+1,n_b+1>+\Psi_{g,0,m}(t)|g,0,m>
$$
Now substitute (\ref{st1}) in (\ref{sh7.1}) and then projecting
the resulting equations onto $<e,n_a,n_b|,$
$<i,n_a,n_b+1|,$$<g,n_a+1,n_b+1|,$ $<g,0,m|$ we can obtain the
following system of equations:
\begin{equation}\label{eq1}
\dot{\Psi}_{e,n_a,n_b}(t)=-i g_be^{-i\delta
\varphi}\sqrt{n_b+1}\Psi_{i,n_a,n_b+1}(t),
\end{equation}
\begin{equation}\label{eq2}
\dot{\Psi}_{i,n_a,n_b+1}(t)=-i g_be^{i\delta
\varphi}\sqrt{n_b+1}\Psi_{e,n_a,n_b}(t)-ig_a\sqrt{n_a+1}\Psi_{g,n_a+1,n_b+1}(t),
\end{equation}
\begin{equation}\label{eq3}
\dot{\Psi}_{g,n_a+1,n_b+1}(t)=-i g_ae^{i\delta
\varphi}\sqrt{n_a+1}\Psi_{i,n_a,n_b+1}(t),
\end{equation}
\begin{equation}\label{eq4}
\dot{\Psi}_{g,0,m}(t)=0.
\end{equation}

To solve the system of equations (\ref{eq1})-(\ref{eq3}) we should
define the initial conditions first. Under initial conditions the
state vector has the following form (see (\ref{5f}))
$$
|\Psi(t=0)>=|\Psi_{atom}\otimes|\Psi_{field}>\equiv
[c_e|e>+c_i|i>+c_g|g>]\otimes\sum_{n_a=0}^{\infty}\sum_{n_b=0}^{\infty}w_{n_a}w_{n_b}|n_a>|n_b>,
$$
or in notations of (\ref{st1})
\begin{equation}\label{ic}
|\Psi(t=0)>=\sum_{n_a,n_b=0}^{\infty}[w_{n_a,n_b}c_e|e,n_a,n_b>+w_{n_a,n_b+1}c_i|i,n_a,n_b+1>+
\end{equation}
$$
+w_{n_a+1,n_b+1}c_g|g,n_a+1,n_b+1>] +w_{0,m}c_g|g,0,m>
$$
where $w_{n_a,n_b}=w_{n_a}w_{n_b}.$

Comparing (\ref{st1}) for $t=0$ and (\ref{ic}) we obtain the initial
conditions
\begin{equation}\label{ic1}
\Psi_{e,n_a,n_b}(t=0)=w_{n_a,n_b}c_e,
\end{equation}
\begin{equation}\label{ic2}
\Psi_{i,n_a,n_b+1}(t=0)=w_{n_a,n_b+1}c_i,
\end{equation}
\begin{equation}\label{ic3}
\Psi_{g,n_a+1,n_b+1}(t=0)=w_{n_a+1,n_b+1}c_g,
\end{equation}
\begin{equation}\label{ic4}
\Psi_{g,0,m}(t=0)=\Psi_{g,0,m}(t=0)=w_{0,m}c_g.
\end{equation}
Now, let us solve the system of equations (\ref{eq1})-(\ref{eq3})
under initial conditions defined by (\ref{ic1})-(\ref{ic4}).
Multiplying (\ref{eq1})-(\ref{eq3}) by $e^{-st}$ and using  the
Laplace transformation of the probability amplitudes the system of
equations transforms to the following:
\begin{equation}\label{sys1}
e^{-st}\dot{\Psi}_{e,n_a,n_b}(t)=-i
g_b\sqrt{n_b+1}e^{-st}e^{-i\delta \varphi}\Psi_{i,n_a,n_b+1}(t),
\end{equation}
\begin{equation}\label{sys2}
e^{-st}\dot{\Psi}_{i,n_a,n_b+1}(t)=-i
g_b\sqrt{n_b+1}e^{-st}e^{i\delta
\varphi}-ig_ae^{-st}\sqrt{n_a+1}\Psi_{g,n_a+1,n_b+1}(t),
\end{equation}
\begin{equation}\label{sys3}
e^{-st}\dot{\Psi}_{e,n_a,n_b}(t)=-i
g_b\sqrt{n_a+1}e^{-st}\Psi_{i,n_a,n_b+1}(t).
\end{equation}

Now  integrating (\ref{sys1})-(\ref{sys3}) by  $t$  we obtain
$$
s\overline{\Psi}_{e,n_a,n_b}(s)+ig_be^{-i\delta
\varphi}\sqrt{n_b+1}\overline{\Psi}_{e,n_a,n_b+1}(s)=\Psi_{e,n_a,n_b}(t=0),
$$
$$
ig_be^{i\delta
\varphi}\sqrt{n_b+1}\overline{\Psi}_{e,n_a,n_b+1}(s)+s\overline{\Psi}_{e,n_a,n_b+1}(s)
+ig_b\sqrt{n_a+1}\overline{\Psi}_{g,n_a+1,n_b+1}(s)=\Psi_{i,n_a,n_b+1}(t=0),
$$
$$
ig_a\sqrt{n_a+1}\overline{\Psi}_{i,n_a,n_b+1}(s)+s\overline{\Psi}_{g,n_a+1,n_b+1}(s)=\Psi_{g,n_a+1,n_b+1}(t=0).
$$
Here $\overline{\Psi}(s)=\int_{0}^{\infty}\Psi(t)e^{-st}dt$ is the
Laplace transform of $\Psi(t).$ This system of equations we also can
write in the following matrix form:
$$
U\left (
\begin{array}{cccc}
\overline{\Psi}_{e,n_a,n_b}(s) \\
\overline{\Psi}_{i,n_a,n_b+1}(s)\\
\overline{\Psi}_{g,n_a+1,n_b+1}(s)
\end{array}
\right)= \left (
\begin{array}{cccc}
w_{n_a,n_b}c_e \\
w_{n_a,n_b+1}c_i\\
w_{n_a+1,n_b+1}c_g
\end{array}
\right),
$$
where
$$
U=\left (
\begin{array}{cccc}
s&ig_ae^{-i\delta\varphi}\sqrt{n_b+1}&0 \\
ig_be^{i\delta\varphi}\sqrt{n_b+1}&s&ig_a\sqrt{n_a+1}\\
0&ig_a\sqrt{n_a+1}&s
\end{array}
\right).
$$

Now if we invoke the inverse Laplace transform we obtain the
following expression for probability amplitudes
\begin{equation}\label{sysf}
\left (
\begin{array}{cccc}
\Psi_{e,n_a,n_b}(t) \\
\Psi_{i,n_a,n_b+1}(t)\\
\Psi_{g,n_a+1,n_b+1}(t)
\end{array}
\right)=\frac{1}{2\pi i} \int_{C}e^{st}U^{-1}(s)ds \left (
\begin{array}{cccc}
w_{n_a,n_b}c_e \\
w_{n_a,n_b+1}c_i\\
w_{n_a+1,n_b+1}c_g
\end{array}
\right),
\end{equation}
where
$$
U^{-1}=\frac{1}{s(s+i\lambda)(1-i\lambda)}\times \left (
\begin{array}{cccc}
s^2+g_a^2(n_a+1)&-isg_ae^{-i\delta\varphi}\sqrt{n_b+1}&-g_be^{-i\delta\varphi}\sqrt{n_b+1}g_a\sqrt{n_a+1} \\
-isg_ae^{-i\delta\varphi}\sqrt{n_b+1}&s^2&-isg_a\sqrt{n_a+1}\\
-g_be^{-i\delta\varphi}\sqrt{n_b+1}g_a\sqrt{n_a+1}&i-isg_a\sqrt{n_a+1}&s^2+g_a^2(n_a+1)
\end{array}
\right),
$$
and $ \lambda=\sqrt{g_a^2(n_a+1)+g_b^2(n_b+1)}.$ The poles of
(\ref{sysf}) coincide with roots of the determinant $U$ and equal
$0,i\lambda,-i\lambda.$ From this, due to the Cauchy second
integral theorem it follows
$$
\Psi_{e,n_a,n_b}(t)=\left\{ \left [ cos(\lambda
t)+\frac{g_a^2}{\lambda^2}(n_a+1)(1-cos(\lambda t))\right
]\Psi_{i,n_a,n_b+1}(0) -\left [ i\frac{g_b}{\lambda}e^{-i\delta
\varphi} \sqrt{n_b+1}sin(\lambda t)\right
]\Psi_{i,n_a,n_b+1}(0)-\right.
$$
$$
\left.-\left[ \frac{g_bg_a}{\lambda^2}e^{-i\delta
\varphi}\sqrt{n_b+1}\sqrt{n_a+1}(1-cos(\lambda t))\right
]\Psi_{g,n_a+1,n_b+1}(0) \right \},
$$
$$
\Psi_{e,n_a,n_b+1}(t)=\left \{ \left [
\frac{g_a}{\lambda}e^{i\delta\varphi}\sqrt{n_b+1}sin(\lambda
t)\right ]\Psi_{e,n_a,n_b}(0) +cos(\lambda t)\right
]\Psi_{i,n_a,n_b+1}(0)-
$$
$$
\left.- \left[ i\frac{g_b}{\lambda}\sqrt{n_a+1}sin(\lambda t)\right
]\Psi_{g,n_a+1,n_b+1}(0) \right\},
$$
$$
\Psi_{e,n_a,n_b}(t)=\left \{ \left[
\frac{g_bg_a}{\lambda_n^2}e^{-i\delta
\varphi}\sqrt{n_b+1}\sqrt{n_a+1}(1-cos(\lambda t))\right ]
\Psi_{i,n_a,n_b}(0) -\left [ i\frac{g_a}{\lambda}
\sqrt{n_a+1}sin(\lambda t)\right ]\Psi_{i,n_a,n_b+1}(0)-\right.
$$
$$
\left.- \left [ cos(\lambda
t)+\frac{g_a^2}{\lambda_n^2}(n_b+1)(1-cos(\lambda t))\right
]\Psi_{g,n_a+1,n_b+1}(0) \right \}.
$$

Amplitudes for the configurations $\Lambda$ and $V$ are obtained
in similar manner.

The evolution operator
$$\left (
\begin{array}{cccc}
u_{11}(t)&u_{12}(t)&u_{13}(t) \\
u_{21}(t)&u_{22}(t)&u_{23}(t) \\
u_{31}(t)&u_{32}(t)&u_{33}(t) \\
\end{array}
\right)
$$
for the $L,\Lambda,V$ configurations has the following form:

$L$-configuration, $ \lambda=\sqrt{g_a^2(n_a+1)+g_b^2(n_b+1)}.$
 $$
 u_{11}^L(t)=cos(\lambda t)+\frac{g_a^2}{\lambda^2}(n_a+1)(1-cos(\lambda t)),
 $$
 $$
u_{12}^L(t)=-i\frac{g_b}{\lambda}e^{-i\delta \varphi}
\sqrt{n_b+1}sin(\lambda t),
 $$
 $$
u_{13}^L(t)= \frac{g_bg_a}{\lambda^2}e^{-i\delta
\varphi}\sqrt{n_b+1}\sqrt{n_a+1}(1-cos(\lambda t)),
 $$
 $$
 u_{21}^L(t)=-i\frac{g_b}{\lambda}e^{i\delta \varphi}
\sqrt{n_b+1}sin(\lambda t),
 $$
 $$
 u_{22}^L(t)=cos(\lambda t),
 $$
 $$
 u_{23}^L(t)=-i\frac{g_b}{\lambda}e^{-i\delta \varphi}
\sqrt{n_a+1}sin(\lambda t),
 $$
 $$
 u_{31}^L(t)=\frac{g_bg_a}{\lambda^2}e^{-i\delta
\varphi}\sqrt{n_b+1}\sqrt{n_a+1}(1-cos(\lambda t)),
 $$
 $$
 u_{32}^L(t)=-i\frac{g_b}{\lambda_n}e^{-i\delta \varphi}
\sqrt{n_a+1}sin(\lambda t),
 $$
 $$
u_{33}^L(t)=cos(\lambda
t)+\frac{g_a^2}{\lambda^2}(n_a+1)(1-cos(\lambda t)).
 $$

$\Lambda$-configuration, $ \lambda=\sqrt{g_a^2(n_a+1)+g_b^2n_b^2}.$
 $$
 u_{11}^{\Lambda}(t)=cos(\lambda t)+\frac{g_a^2}{\lambda^2}(n_a+1)(1-cos(\lambda t)),
 $$
 $$
u_{12}^{\Lambda}(t)=i\frac{g_b}{\lambda}e^{-i\delta \varphi}
\sqrt{n_b}sin(\lambda t),
 $$
 $$
u_{13}^{\Lambda}(t)= \frac{g_bg_a}{\lambda^2}e^{-i\delta
\varphi}\sqrt{n_b}\sqrt{n_a+1}(1-cos(\lambda t)),
 $$
 $$
 u_{21}^{\Lambda}(t)=i\frac{g_b}{\lambda}e^{i\delta \varphi}
\sqrt{n_b}sin(\lambda t),
 $$
 $$
 u_{22}^{\Lambda}(t)=cos(\lambda t),
 $$
 $$
 u_{23}^{\Lambda}(t)=-i\frac{g_b}{\lambda}
\sqrt{n_a+1}sin(\lambda t),
 $$
 $$
 u_{31}^{\Lambda}(t)=\frac{g_bg_a}{\lambda^2}e^{i\delta
\varphi}\sqrt{n_b}\sqrt{n_a+1}(1-cos(\lambda t)),
 $$
 $$
 u_{32}^{\Lambda}(t)=-i\frac{g_a}{\lambda}\sqrt{n_a+1}sin(\lambda t),
 $$
 $$
u_{33}^{\Lambda}(t)=cos(\lambda
t)+\frac{g_b^2}{\lambda^2}n_b(1-cos(\lambda t)).
 $$

V-configuration, $ \lambda=\sqrt{g_a^2n_a+g_b^2(n_b+1)}.$
$$
 u_{11}^V(t)=cos(\lambda t)+\frac{g_a^2}{\lambda^2}(n_a)(1-cos(\lambda t)),
 $$
 $$
u_{12}^V(t)=-i\frac{g_b}{\lambda}e^{-i\delta \varphi}
\sqrt{n_b+1}sin(\lambda t),
 $$
 $$
u_{13}^V(t)= \frac{g_bg_a}{\lambda^2}e^{-i\delta
\varphi}\sqrt{n_b+1}\sqrt{n_a}(1-cos(\lambda t)),
 $$
 $$
 u_{21}^V(t)=-i\frac{g_b}{\lambda}e^{i\delta \varphi}
\sqrt{n_b+1}sin(\lambda t),
 $$
 $$
 u_{22}^V(t)=cos(\lambda t),
 $$
 $$
 u_{23}^V(t)=i\frac{g_a}{\lambda}
\sqrt{n_a}sin(\lambda t),
 $$
 $$
 u_{31}^V(t)=\frac{g_bg_a}{\lambda^2}e^{i\delta
\varphi}\sqrt{n_b+1}\sqrt{n_a}(1-cos(\lambda t)),
 $$
 $$
 u_{32}^V(t)=i\frac{g_a}{\lambda}\sqrt{n_a}sin(\lambda t),
 $$
 $$
u_{33}^V(t)=cos(\lambda
t)+\frac{g_b^2}{\lambda^2}(n_b+1)(1-cos(\lambda t)).
 $$

Under the assumptions described above we have constructed
evolution operators in explicit form. Performed analysis has shown
that unlike the case of methods of semiclassical analysis ---
where the atom is quantized whereas the field is classical, the
completely quantum approach shows that in the absence of fields
the atom still oscillates and this must be taken into account in
quantum calculations. Moreover quantum processing may be performed
on the level of single photons.

\end{document}